\documentclass[fleqn,usenatbib]{mnras}

\usepackage[T1]{fontenc}
\usepackage{xcolor}

\DeclareRobustCommand{\VAN}[3]{#2}
\let\VANthebibliography\thebibliography
\def\thebibliography{\DeclareRobustCommand{\VAN}[3]{##3}\VANthebibliography}

\usepackage{graphicx}	
\usepackage{amsmath}	
\usepackage{amssymb}	
\usepackage{lipsum}
\usepackage{xcolor}
\usepackage{newtxtext,newtxmath,widetext}

\title[Multimessenger time-domain outlier analysis]{Controlling Outlier Contamination In Multimessenger Time-domain Searches For Supermasssive Binary Black Holes}

\author[Wang \& Taylor]{
Qiaohong Wang,$^{1}$\thanks{E-mail: qiaohong.wang@vanderbilt.edu}
Stephen~R.~Taylor$^{1}$\thanks{E-mail: stephen.r.taylor@vanderbilt.edu}
\\
$^{1}$Department of Physics \& Astronomy, Vanderbilt University, 2301 Vanderbilt Place, Nashville, TN 37235, USA\\
}

\date{Accepted XXX. Received YYY; in original form ZZZ}

\begin{document}
\label{firstpage}
\pagerange{\pageref{firstpage}--\pageref{lastpage}}
\maketitle

\begin{abstract}
Time-domain datasets of many varieties can be prone to statistical outliers that result from instrumental or astrophysical anomalies. These can impair searches for signals within the time series and lead to biased parameter estimation. Versatile outlier mitigation methods tuned toward multimessenger time-domain searches for supermassive binary black holes have yet to be fully explored. In an effort to perform robust outlier isolation with low computational costs, we propose a Gibbs sampling scheme. This provides structural simplicity to outlier modeling and isolation, as it requires minimal modifications to adapt to time-domain modeling scenarios with pulsar-timing array or photometric data. We robustly diagnose outliers present in simulated pulsar-timing datasets, and then further apply our methods to pulsar J$1909$$-$$3744$ from the NANOGrav 9-yr Dataset. We also explore the periodic binary-AGN candidate PG$1302$$-$$102$ using datasets from the Catalina Real-time Transient Survey, All-Sky Automated Survey for Supernovae, and the Lincoln Near-Earth Asteroid Research. We present our findings and outline future work that could improve outlier modeling and isolation for multimessenger time-domain searches.
\end{abstract}

\begin{keywords}
galaxies: active -- (galaxies:) quasars: supermassive black holes -- gravitational waves -- (stars:) pulsars: general -- methods: data analysis
\end{keywords}



\section{Introduction}

Supermassive black holes (SMBHs) are believed to reside at the center of most massive galaxies \citep[e.g.,][]{2013ARA&A..51..511K,HaehneltKauffmann2002}, with the formation of SMBH binaries (SMBHBs) occurring as a natural by-product within the current hierarchical galaxy-growth paradigm \citep{1980Natur.287..307B}.
Despite significant efforts, unambiguous evidence of SMBHBs at sub-parsec separations remains elusive. Yet once found,  they will significantly constrain models of galaxy formation and evolution \citep[e.g.,][]{Hopkins_2005a,Hopkins_2005b}, massive black-hole growth, and the final-parsec dynamical interactions of massive black holes before coalescence \citep{2001ApJ...563...34M,yu2002,Colpi_2014,Khan_2016,2018MNRAS.473.3410R,Goicovic2018AccretionOC,Mu_oz_2020,2017PhRvL.118r1102T,Chen_2019}.

In general, the coalescence of two SMBHs after a galaxy merger can be categorized into three stages: $(i)$ a dynamical friction stage, where the cores of the merging galaxies sink within the common remnant, bringing the SMBHs to within $\sim 1$~pc of each other; $(ii)$ binary hardening via stellar loss-cone scattering, circumbinary disk interactions, and three-body encounters in the sub-parsec regime; and $(iii)$ binary inspiral through gravitational wave (GW) emission in the $\sim$~milli-parsec regime. 
Throughout these stages, both widely-separated pairs and bound systems may produce a variety of observable signatures. 
When the separation of the pair of SMBHs is of order  kiloparsecs, the system can be observed as a dual active galactic nucleus (AGN) \citep[e.g.,][]{Comerford_2015}. As the system's separation tightens, it becomes more challenging to individually resolve both radio cores. The smallest separation system that has been directly resolved into its two components is $\sim 7$~pc \citep{2006ApJ...646...49R}. At smaller separations, particularly the milli-parsec regime where GW emission may be detectable, more indirect methods are used to attempt to find SMBHBs. 

Pulsar Timing Arrays (PTAs) search for $\sim1-100$~nHz-frequency GW emission from milli-parsec--separated SMBHBs, where the measurable influence corresponds to correlated timing deviations across widely separated pulsars in the Milky Way. Current leading organizations include the North American Nanohertz Observatory for Gravitational waves (NANOGrav, \citet{2019BAAS...51g.195R}), the European Pulsar Timing Array (EPTA, \citet{Desvignes_2016}), the Parkes Pulsar Timing Array (PPTA, \citet{kerr_2020}), and the Indian Pulsar Timing Array (InPTA), all of which constitute the International Pulsar Timing Array (IPTA, \citet{Perera_2019}). The IPTA (and regional PTAs that comprise it), as well as other emerging PTAs and radio-telescope programs (e.g., the Chinese PTA, CHIME/PTA, MeerTIME, etc.) are expected to detect a stochastic GW background within the next several years, and individual binaries by the end of the 2020s \citep{2016ApJ...819L...6T,2021ApJ...911L..34P,2017MNRAS.471.4508K,2018MNRAS.477..964K,2017NatAs...1..886M}. While no GW detection has yet been made, PTAs have placed informative constraints on the presence and properties of SMBHBs within the local Universe. For example, the NANOGrav $11$~year dataset was recently used to constrain the mass ratio of putative binaries in $216$ local galaxies \citep{Arzoumanian_2021}. 

Significant progress also has been made in indirectly identifying SMBHBs using large optical time-domain surveys that search for periodically varying quasars. \citet{Graham_2015b,Graham_2015a} reported $111$ binary black hole candidates from $243500$ qussar optical variability analysis based on the Cataline Real-time Transient Survey (CRTS). \citet{2016Charisi} identified $33$ binary candidates out of $35383$ qusars periocity searches in the Palomar Transient Factor (PTF) data. \citet{Liu_2015} claimed one periodic candidate using photometric data from the Pan-STARRS1 Medium Deep Survey, which has since proven to be insignificant in an extended analysis \citep{Liu_2016}. \citet{Liu_2019} further searched over $\sim 9000$ quasars and reported one candidate. \citet{Chen_2020} reported five candidates from a systematic searches of Dark Energy Survey Supernova (DES-SN) fields and the Sloan Digital Sky Survey Stripe 82 (SDSS-S82). Though many SMBHB candiates are reported, their validity remains uncertain \citep{2016MNRAS.461.3145V,2018ApJ...856...42S,Zhu_2020}. 

Several factors impose challenges to identifying SMBHBs in time-series analyses, including, e.g., data quality, noise, and insufficient dataset length. Due to the stochastic behavior of normal quasar variability, it is possible that individual quasars can be mis-identified as periodic binary candidates \citep{2016MNRAS.461.3145V}. Even if a binary were present, intrinsic quasar variability renders it more challenging to tease out underlying periodic signals, and frequent false-positive detections are expected \citep{2021arXiv211007465W}. Similar challenges exist in pulsar-timing analyses, where pulsar intrinsic red noise can be conflated with a periodic signature or the statistical behavior expected from a GW background. Some solutions are to harness data over a longer period of time to discern the statistical difference between random noise and a deterministic periodic signal, and to improve the overall quality of the measurements. However, limitations on survey time-spans and telescope technology inhibit what can be expected in the near term. Hence we turn our focus to robust and powerful statistical methods that can be readily deployed to isolate the influence of outlier contamination on the target model space, and thereby improve the overall quality of the dataset for multi-messenger searches. 

Most current systematic searches for SMBHBs using AGN lightcurves involve applying filters, fitting the filtered lightcurve, then clipping points that lie significantly far from the fit (so-called ``sigma clipping''). \citet{Graham_2015a,Graham_2015b} apply a $3$-point median filter, and remove points that deviate from a quintic polynomial fit to the data with a clipping threshold of 0.25 mag.  \citet{Liu_2016} correlates the magnitude with the standard deviation, fit the binned relation to a parabola and excludes points at $4.5\sigma$. \citet{2016Charisi} apply a $3$-point median filter and exclude points that deviate by three sigma or more from a quintic polynomial fit. Lastly, \citet{Chen_2020} first clips data with $5\sigma$ from the median in each band and further remove the data points with $3\sigma$ after combining light curves from different telescopes. While implementing sigma clipping makes intuitive sense, it differs from survey to survey, and it is challenging to generalize such method to other datasets, especially for combined multimessenger analysis. In NANOGrav, the preparation of pulsar-timing datasets involves a Bayesian outlier step, which statistically assigns probabilities of each data point being an outlier from the ``inlier'' model \citep{Vallisneri_2017}. Expanding this kind of Bayesian outlier modeling to photometric lightcurves would provide a new way to statistically boost the data quality, and offer a new outlook for multimessenger searches for SMBHBs \citep{2022MNRAS.510.5929C}.

The rest of the paper is laid out as follows: in Section~\ref{sec:methods} we introduce our outlier modeling framework, including an efficient Gibbs scheme to sample the parameter space, and a simple sinusoidal time-series to showcase the implementation. In Section~\ref{sec:results} we apply the proposed outlier model to two time-domain datasets: PTA data and photometric AGN lightcurves. We illustrate the adaptation and implementation of our outlier isolation method to simulated PTA data, and real pulsar J$1909$$-$$3744$ data derived from the \textit{NANOGrav 9-year Dataset}. 
We further adapt our outlier modeling to photometric AGN lightcurves, and deploy it on the binary-AGN candidate PG$1302$$-$$102$ using datasets from CRTS, the All-Sky Automated Survey for Supernovae (ASAS-SN), and the Lincoln Near-Earth Asteroid Research (LINEAR). Finally, in Section~\ref{sec:conclusions} we discuss our findings and future work that could improve outlier modeling and isolation for multimessenger time-domain searches.
 
\section{Methods}\label{sec:methods}

\subsection{An outlier model}

An outlier is defined to be an anomalous event or observation that arises from a process that differs from the majority of the data generation.  
Outlier detection has always been an indispensable part of data analysis; contaminated datasets without proper outlier treatment can increase modeling uncertainties and produce misleading results. A common mitigation strategy is to model the data as a Gaussian mixture of inlier and outlier distributions with differing variance \citep{hogg2010data,Vallisneri_2017}.    
For example, assume a model with an inlier data model, $y_i = \mu_i + \epsilon_i$, where $y_i$ is the $i$-th observation, $\mu_{i}$ is the mean, and $\epsilon_i$ is Gaussian random noise with variance $\sigma_i^2$. We can then account for possible outlier contamination in the data collection by modeling such fluctuations with $\epsilon_\mathrm{out}\sim\mathcal{N}(0, \sigma_\mathrm{out}^{2})$, where our full data model now includes a latent outlier indicator $z_i$ for each observation, where $z_i=1$ for an outlier and $0$ otherwise. This indicator is modeled with a certain prior outlier probability, $\theta$, e.g., $z_i \sim \mathrm{Bernoulli}(\theta)$. This data model is a mixture of two Gaussians with different mean and different variances, which can be expressed in the form
\begin{equation} \label{eq:outlier}
y_i = (1-z_i) \mu_i +(1-z_i) \epsilon_i + z_i \epsilon_\mathrm{out}.
\end{equation}
Incorporating the outlier indication parameter $z_i$ allows us to assess the possibility of an individual measurement being an outlier, and to formulate the likelihood as a mixture of two Gaussians that respectively model the inlier and outlier distributions: 
\begin{equation} \label{eq:out_model}
    p \left(y_{i} \mid \mu_{i}, z_{i}, \sigma_{i}, \sigma_{\mathrm{out }}\right)=\left\{\begin{array}{ll}
    \mathrm{e}^{-\left(y_{i}- \mu_{i}\right)^{2} /\left(2 \sigma_{i}^{2}\right)} / \sqrt{2 \pi \sigma_{i}^{2}}, & \text{for } z_{i}=0 \\
    \mathrm{e}^{-y_{i}^{2} /\left(2 \sigma_{\mathrm{out }}^{2}\right)} / \sqrt{2 \pi \sigma_{\mathrm{out }}^{2}}, & \text{for } z_{i}=1
    \end{array}\right.
\end{equation}
We use this model for the toy sine wave example in Sec.~\ref{sec:sinewave}.

\citet{Ellis_2019} introduced a more flexible approach that generalizes the above outlier modeling through an auxiliary variable $\alpha_i$, which allows each outlier point to be drawn from a Gaussian distribution whose variance is a multiple of the inlier model. Upon incorporating this auxiliary variable, \citet{Ellis_2019} suggest a data model that is a mixture of Gaussians with the same mean and different variances:
\begin{align}\label{eq:out_model2}
    \epsilon_i \mid z_i, \alpha_i &\sim \mathcal{N}\left(0, \alpha_i^{z_i} \sigma_i^2\right), \nonumber\\
    z_i \mid \theta &\sim \mathrm{Bernoulli}(\theta) \nonumber\\
    \theta &\sim \mathrm{Beta} / \mathrm{Gamma} / \mathrm{Normal} \nonumber\\
    \alpha_i \mid \nu &\sim \mathrm{InverseGamma}(\nu / 2, \nu / 2) \nonumber\\
    \nu &\sim \mathrm{Uniform}(1,40).
\end{align}

The choice of prior outlier probability, $\theta$, can vary depending on the sample size, sampling scheme, and assumed outlier proportion (i.e., confidence in the underlying data quality). See \citet{Tak_2017} and \citet{Ellis_2019} for a more thorough discussion on the proper choice of prior. By including $\alpha_i$ as a scaling factor to the data variance, this approach can easily model each data point as either an inlier or outlier depending on the latent indicator variable, $z_i$. Furthermore, by phrasing the data model as a single Gaussian distribution (albeit with hyper-parameters), this approach minimizes disruption to the original ``outlier-less'' model structure, thereby endowing it with versatility for a variety of hierarchical modeling scenarios. We use this model for the pulsar timing and photometric lightcurve scenarios in Sec.~\ref{sec:results}. We note that if we set $\alpha$ to be constant, the data model is a mixture of two Gaussian distributions with the same mean and different variances. Yet if we are actively searching for $\alpha$, the data model becomes a mixture of a Gaussian distribution and a Student-$t$ distribution \citep{Ellis_2019}.

\subsection{Gibbs sampling}

Of primary interest in performing outlier analyses of a dataset is the construction of posterior marginal distributions of the outlier probability for each observation. This is a challenging endeavor, especially as datasets grow in volume. 
Recent work in PTA analysis has deployed hierarchical Bayesian inference of the outlier model space using a Hamiltonian Monte Carlo scheme, which is advantageous for exploring large hierarchical model spaces \citep{Vallisneri_2017}. However, to apply this sampling technique, one needs to compute the gradient of the target probability density function that defines a vector field in parameter space \citep{Neal2011,betancourt2018conceptual}. The difficulty in obtaining differential equations makes Hamiltonian Monte Carlo challenging to be generalized to varying data structures. 

An alternative approach is available through Gibbs sampling, which provides the ability to carry out a fully Bayesian outlier analysis with relative simplicity. The essence of Gibbs lies in directly sampling from each parameter's (or parameter blocks') conditional posterior distribution iteratively \citep{GelfandSmith1990}. This contrasts with the operations of more common acceptance-rejection sampling techniques, which propose new sample positions for all parameters simultaneously, and which are subsequently assessed by a metric (e.g., the Metropolis-Hastings ratio) to determine whether the new parameter vector is accepted. There are significant benefits to Gibbs sampling: $(i)$ by drawing samples directly from the conditional posteriors, the MCMC chain typically has lower autocorrelation length than in acceptance-rejection analyses; $(ii)$ by not rejecting any points, Gibbs sampling is faster at generating a required set of random samples from a target probability distribution. The key drawback of Gibbs sampling is that one must know the form of conditional parameter posteriors, and how to directly sample from it. 

For a generic problem with a likelihood and parameter priors, the conditional posteriors will not necessarily have the form of standard probability distributions. However, we can partially remedy this through the use of \textit{conjugate priors}, such that the conditional posterior lies in the same family of distributions as the prior. This can provide a closed-form solution to the conditional posterior distribution that facilitates directly sampling from standard distribution families, rather than acceptance-rejection schemes. For example, if a problem calls for a Gaussian likelihood function with mean and variance parameters, the conjugate prior for fixed variance is a Gaussian distribution on the mean. This leads to a Gaussian conditional posterior for the mean given the data and fixed variance. By contrast, the conjugate prior for fixed mean is an inverse Gamma distribution on the variance. Finding appropriate conjugate priors is not always possible, and when the conditional posterior is a non-standard distribution alternative approaches can be used, e.g., weighted bootstrap \citep{SmithGelfand1992_weighted}, approximate cdf inversion \citep{RitterTanner1992}, and conventional Metropolis-Hastings sampling \citep{ChibGreenberg}. The latter approach is used here for parameters that don't have conjugate priors, where we replace a step involving a direct conditional draw with a short Metropolis-Hastings MCMC that returns a single random sample after a sufficient number of iterations \citep{Gelfand2000}. 

The general process for Gibbs sampling involves initializing all parameters with some values, then sequentially updating parameters (either through draws from conditional distributions or short MCMC chains). As each parameter is updated, the other parameters in the model are fixed to their most recently updated values. Once all parameters have been cycled through, the process is iterated as appropriate. This procedure is shown in the following algorithm for a model with $n_\mathrm{dim}$ parameters $\{X_1,X_2,\ldots,X_{n_\mathrm{dim}}\}$:
\begin{equation}
\begin{array}{l}
    \text { 1: } \text { Initialization } x^{(0)} \sim p(x) \\
    \text { 2: } \text { for } \{i=1,2, \ldots N\} \text { do } \\
    \text { 3: } \quad x_{1}^{(i)} \sim p\left(X_{1}=x_{1} \mid X_{2}=x_{2}^{(i-1)}, X_{3}=x_{3}^{(i-1)}, \ldots, X_{n_{\text {dim }}}=x_{n_{\text {dim }}}^{(i-1)}\right) \\
    \text { 4: } \quad x_{2}^{(i)} \sim p\left(X_{2}=x_{2} \mid X_{1}=x_{1}^{(i)}, X_{3}=x_{3}^{(i-1)}, \ldots, X_{n_{\operatorname{dim}}}=x_{n_\mathrm{dim}}^{(i-1)}\right) \\
    5: \quad \vdots \\
    \text { 6: } \quad x_{n_{\text {dim }}}^{(i)} \sim p\left(X_{n_{\text {dim }}}=x_{n_{\text {dim }}} \mid X_{1}=x_{1}^{(i)}, X_{2}=x_{2}^{(i)}, X_{3}=x_{3}^{(i)}, \ldots\right) \\
    \text { 7: } \quad i=i+1 \\
    \text { 8: end for }
\end{array}
\end{equation}

In Sec.~\ref{sec:sinewave}, we perform a simple demonstration of our outlier modeling scheme with Gibbs sampling on a toy model with data from an underlying sine wave, with noise and outliers.

\begin{table*}
\begin{center}
\renewcommand{\arraystretch}{1.3}
\setlength{\tabcolsep}{6.5pt}
\small
\begin{tabular}{llll}
\hline\hline
\multicolumn{1}{c}{\bf{Parameter}}&
\multicolumn{1}{c}{\bf{Description}} &
\multicolumn{1}{c}{\bf{Prior}} &
\multicolumn{1}{c}{\bf{Prior Parameters}}\\ 
\hline
\multicolumn{4}{c}{\bf{Sinusoidal wave}} \\[1pt]
$a$ & amplitude & $\mathcal{N}[a_{0}, \sigma_{0}^{2}]$ & $a_0=5, \sigma_0=2$  \\
$\sigma_{i}$ & variance on observation & InverseGamma $[v_{0}, \beta_{0}]$& $\nu_0=1, \beta_0=3$ \\
$P$ [days] & period & Uniform $[P_{\mathrm{min}},P_{\mathrm{max}}]$& $P_\mathrm{min}=5$~days, $P_\mathrm{max}=30$~days \\
$z_{i}$ & outlier tag on observation & Bernoulli $[\theta]$& $\theta = 0.1$\\
$\theta$ & outlier probability & Beta $[\alpha,\beta]$ & $\alpha=1, \beta=15$\\
\hline

\multicolumn{4}{c}{\bf{Pulsar timing}} \\[1pt]
$E_{k}$ & EFAC per backend/receiver system & Uniform $[E_{k,\mathrm{min}}, E_{k,\mathrm{max}}]$ & $E_{k,\mathrm{min}}=0.1, E_{k,\mathrm{max}}=10$  \\
$Q_{k}$ [s] & EQUAD per backend/receiver system & log$_{10}$-Uniform $[Q_{k,\mathrm{min}}, Q_{k,\mathrm{max}}]$ &$Q_{k,\mathrm{min}} =-10,Q_{k,\mathrm{max}}= -4$ \\
$J_{k}$ [s] & ECORR per backend/receiver system & log$_{10}$-Uniform $[J_{k,\mathrm{min}}, J_{k,\mathrm{max}}]$ & $J_{k,\mathrm{min}} =-10,J_{k,\mathrm{max}}= -4$ \\
$A_{\rm red}$ & red-noise power-law amplitude &log$_{10}$-Uniform$[A_{\rm red, \mathrm{min}},A_{\rm red, \mathrm{max}}]$&$ A_{\rm red, \mathrm{min}}=-18, A_{\rm red, \mathrm{max}}-11$  \\
$\gamma_{\rm red}$ & red-noise power-law spectral index & Uniform $[\gamma_{\rm red, \mathrm{min}},\gamma_{\rm red, \mathrm{max}}] $&$\gamma_{\rm red, \mathrm{min}}=0, \gamma_{\rm red, \mathrm{min}}=7$  \\
$z_{i}$ & outlier tag on observation & Bernoulli$[\theta]$&$\theta = 0.01$\\ 
$\theta$ & outlier probability & Beta$[km,k(1-m)]$ & $k=N_\mathrm{data}, m=0.01$ \\ 
$\alpha_{i}$ & scaling factor of variance & InverseGamma$(\nu / 2, \nu / 2)$ & $\nu=4$ \\
\hline

\multicolumn{4}{c}{\bf{Photometric lightcurves}} \\[1pt]
$m$ [mag] & mean magnitude with offsets & Uniform $[m_\mathrm{min},m_\mathrm{max}]$& $m_\mathrm{min}=14.5$~mag, $m_\mathrm{max}=15.5$~mag  \\
$A$ [mag] & amplitude of a periodic signal $\vartheta_{s}$& Uniform $[A_\mathrm{min},A_\mathrm{max}]$ &$A_\mathrm{min}=0$~mag, $A_\mathrm{max}=0.5$~mag  \\
$\phi$ & phase angle of a periodic signal $\vartheta_{s}$ & Uniform $[\phi_\mathrm{min},\phi_\mathrm{max}]$& $\phi_\mathrm{min}=0, \phi_\mathrm{max}= 2\pi$  \\
$1/f_{0}$ [years] & period of a periodic signal $\vartheta_{s}$ & Uniform $[(1/f_{0})_\mathrm{min},(1/f_{0})_\mathrm{max}]$ & $(1/f_{0})_\mathrm{min}= 0$~years, $(1/f_{0})_\mathrm{max}=10$~years  \\
$\gamma$ &red-noise power-law spectral index& Uniform$[\gamma_\mathrm{min},\gamma_\mathrm{max}]$ & $\gamma_\mathrm{min}=0, \gamma_\mathrm{max}=1.8$\\
$E$ & EFAC per observation & Uniform $[E_\mathrm{min},E_\mathrm{max}]$&$E_\mathrm{min}=0.1,E_\mathrm{max}=2.0$\\
$\hat{\sigma}^{2}$ [mag$^2$yr$^{-1}$] &intrinsic variance
between observations& ln-Uniform $[\hat{\sigma}^{2}_\mathrm{min}, \hat{\sigma}^{2}_\mathrm{max}]$&$ \hat{\sigma}^{2}_\mathrm{min}=-6, \hat{\sigma}^{2}_\mathrm{max}=0$ \\
$\tau_{0}$ [years] & noise
damping timescale& ln-Uniform $[\tau_{0,\mathrm{min}}, \tau_{0,\mathrm{max}}]$&$\tau_{0,\mathrm{min}}=-4, \tau_{0,\mathrm{max}}=4$\\
$z_{i}$ & outlier tag on observation & Bernoulli$[\theta]$& $\theta = 0.01$\\
$\theta$ & outlier probability & Beta $[km,k(1-m)]$ & $k=N_\mathrm{data}, m=0.01$\\ 
$\alpha_{i}$ & scaling factor of variance & InverseGamma$(\nu / 2, \nu / 2)$ & $\nu=4$ \\
\hline

\end{tabular}
\caption{Prior distributions for the model parameters in all of our analyses. The sinusoidal wave corresponds to Sec.~\ref{sec:sinewave}. The pulsar timing case corresponds to Sec.~\ref{sec:pulsar_timing}. The photometric lightcurve case corresponds to Sec.~\ref{sec:photo_data}. In cases where the prior is labeled as log$_{10}$-Uniform (ln-Uniform), this means that the prior is uniform in the base-$10$ logarithm (natural logarithm) of the parameter, with ranges on the logarithm indicated in the rightmost column. \label{tab:priors}}
\end{center}
\end{table*}

\subsection{Toy sine wave} \label{sec:sinewave}

Sinusoidal models are commonly used to search for periodic structure in time-domain data. The following example serves as a practical guide to the Gibbs outlier scheme that we will later deploy on pulsar-timing and photometric lightcurve data. Our data model is given by $y_i = a \sin(2\pi t_i/P) + \epsilon_i$, corresponding to an underlying sine-wave signal model with Gaussian random noise added. We also add outlier data, $\epsilon_\mathrm{out}$, that is drawn from a very different (zero mean) statistical distribution: $\epsilon_\mathrm{out}\sim\mathcal{N}(0, \sigma_\mathrm{out}^{2})$. This emulates the scenario of outlier observations being caused by instrumental artifacts that may not follow the sinusoidal behavior of the signal. As discussed in \citet{Vallisneri_2017}, strong thermal-noise spikes in radio instrumentation can create outliers in low-SNR pulsar timing observations; not only are the outliers spread much more broadly than
inlier data, but they do not center around the true observations. 
 
\begin{figure}
	\includegraphics[width=\columnwidth]{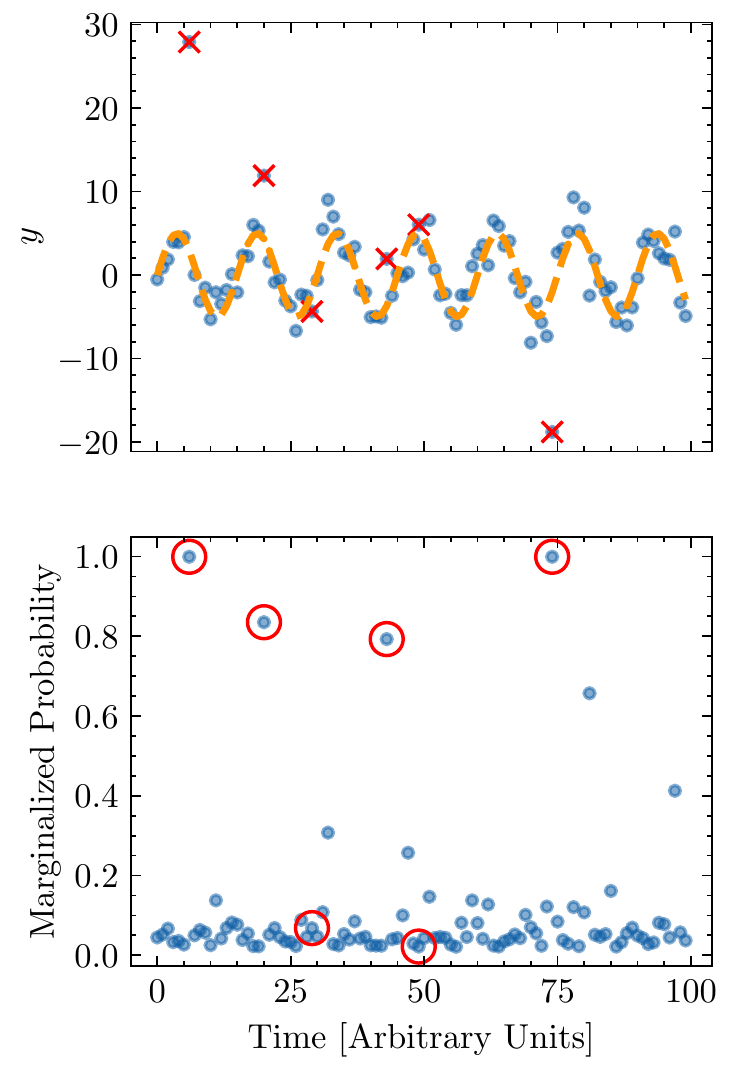}
    \caption{\textit{Top:} simulated noisy sine wave data points (blue dots) with outliers marked in red, overlayed with the injected sine wave (orange dotted line). \textit{Bottom:} marginalized probability for each datapoint being an outlier, true outliers circled in red.}
    \label{fig:picture1}
\end{figure}

In our numerical example, we generate 100 equally-spaced data points from a sine wave model with $a=5$, $P =15$, $\sigma_{i}=2$, with a $5\%$ probability for points to be drawn from an outlier Gaussian distribution $\mathcal{N}(0, 20)$. This toy dataset is shown in the top panel of \autoref{fig:picture1}. To illustrate how our method can discriminate data points drawn from the true underlying signal model versus outliers, we deploy the model in \autoref{eq:outlier}  and \autoref{eq:out_model}. By expanding the data model to include outlier tags for each data point (and associated hyperparameters) we get the following structure for our hierarchical model:

\begin{align}
    y_{i} \mid a,\sigma_{i}^{2}, P, z_{i} &\sim \mathcal{N} \left((1-z_i)\: a \sin(2\pi t_i/P),(1-z_i)\:\sigma_{i}^{2} + z_i\:\sigma_\mathrm{out}^{2}\right) \nonumber\\ 
    a &\sim \mathcal{N}(a_{0}, \sigma_{0}^{2}) \nonumber\\
    \sigma_{i}^{2} &\sim \mathrm{InverseGamma}\left(v_{0}, \beta_{0}\right), \nonumber\\
    P &\sim \text {Uniform}(P_\mathrm{min},P_\mathrm{max}) \nonumber\\
    z_{i} &\sim \text {Bernoulli}(\theta), \nonumber\\
    \theta &\sim \text {Beta}(\alpha, \beta)
\end{align}

In order to implement this outlier model with a Gibbs sampler, we have selectively chosen conjugate priors for the parameters $a$ (amplitude), $\sigma_i$ (variance on observation), $z_i$ (outlier tag on observation), and $\theta$ (prior outlier probability), which have the hyperparameters $\{a_{0},\sigma_{0}^{2},v_{0}, \beta_{0},\alpha, \beta\}$. The choices of these prior hyperparameters for this and other cases in this paper are shown in \autoref{tab:priors}. This allows the conditional posterior distributions for these parameters to be written in the form of standard families of probability distributions. There is no convenient conditional distribution for the period, thus we implement a Metropolis-Hastings acceptance-rejection step for $P$ with a uniform prior distribution between $5$ and $30$ units. Additionally, after each iteration we define $\{y_i^{*}=(1-z_i)y_i,t^{*}=(1-z_i)t_i\}$ based on the current $z_i$, which corresponds to selecting a subset of the original data where the outliers have been momentarily eliminated \citep{Verdinelli1991BayesianAO}. Following \autoref{eq:out_model2}, and letting $\boldsymbol{\eta}$ denote all other parameters, we update each variable as follows:

\begin{widetext}
\begin{align} \label{euqation6}
    \pi_{1}(a \mid  \boldsymbol{\eta},y_{i},z_i) &\sim \mathcal{N}\left(\frac{\sum_{i=1}^{n} y_{i}^{*} \sin (2 \pi t_{i}^{*}/P)/\sigma^{2}_i + a_0 / \sigma_0^2}{\sum_{i=1}^{n}\sin^{2}(2 \pi t_{i}^{*}/P) / \sigma^{2}_{i}+1 / \sigma_{0}^{2}}, \left[\sum_{i=1}^{n}\sin^{2}(2\pi t_{i}^{*} / P)/\sigma_{i}^{2} + 1/\sigma_{0}^{2}\right]^{-1}\right) \nonumber\\
    \pi_{2}(\sigma_{i}^{2} \mid \boldsymbol{\eta},y_{i},z_i) &\sim \mathrm{InverseGamma}\left(\nu_{0}+\frac{n}{2}, \beta_{0}+\frac{1}{2}\sum_{i=1}^{n}\left[y_{i}^{*}-a\sin(2\pi t_{i}^{*}/P)\right]^{2}\right) \nonumber\\
    \pi_{3}(P\mid \boldsymbol{\eta},y_{i},z_i) &\sim \mathrm{MetropolisHastings}\nonumber\\
    \pi_{4}(z_{i} \mid \boldsymbol{\eta},y_i) &\sim \operatorname{Bernoulli}\left(\frac{\theta\times \mathcal{N}(y_i\mid 0,  \sigma_\mathrm{out}^{2})}{(1-\theta)\times\mathcal{N}(y_i\mid a\sin(2\pi t_i/P),  \sigma_{i}^{2}) + \theta\times \mathcal{N}(y_i\mid 0,  \sigma_\mathrm{out}^{2})}\right)\nonumber\\
    \pi_{5}(\theta \mid \boldsymbol{\eta},y_i) &\sim \operatorname{Beta}\left(\alpha+\sum_{i=1}^{n} z_{i}, n+\beta-\sum_{i=1}^{n} z_{i}\right)
\end{align}
\end{widetext}

We implement the Gibbs sampling scheme by cycling through the set of conditional distributions and drawing one sample from each. After each draw or each stage, the succeeding distributions are updated with the new value of each parameter. 
Sampling is initialized with the following parameter values: $a = 4$, $P = 10$, $\sigma_{i}^2=3$, $\theta = 0.1$. Hyperparameter values are chosen based on empirical tests, and are set to the following values: $a_{0} = 5$, $\sigma_{0}^2 = 4$, $\nu_{0} = 1$, $\beta_{0}= 3$, $\alpha = 1$, $\beta = 15$. In the first Gibbs iteration, we run the Metropolis-Hasting MCMC chain on the period for $2000$ steps in order to burn-in and get closer to the target value. For the rest of the $10^4$ Gibbs iterations, we only perform a short $20$-step Metropolis-Hastings chain in each for the period.

\begin{figure}
    \includegraphics[width=\columnwidth]{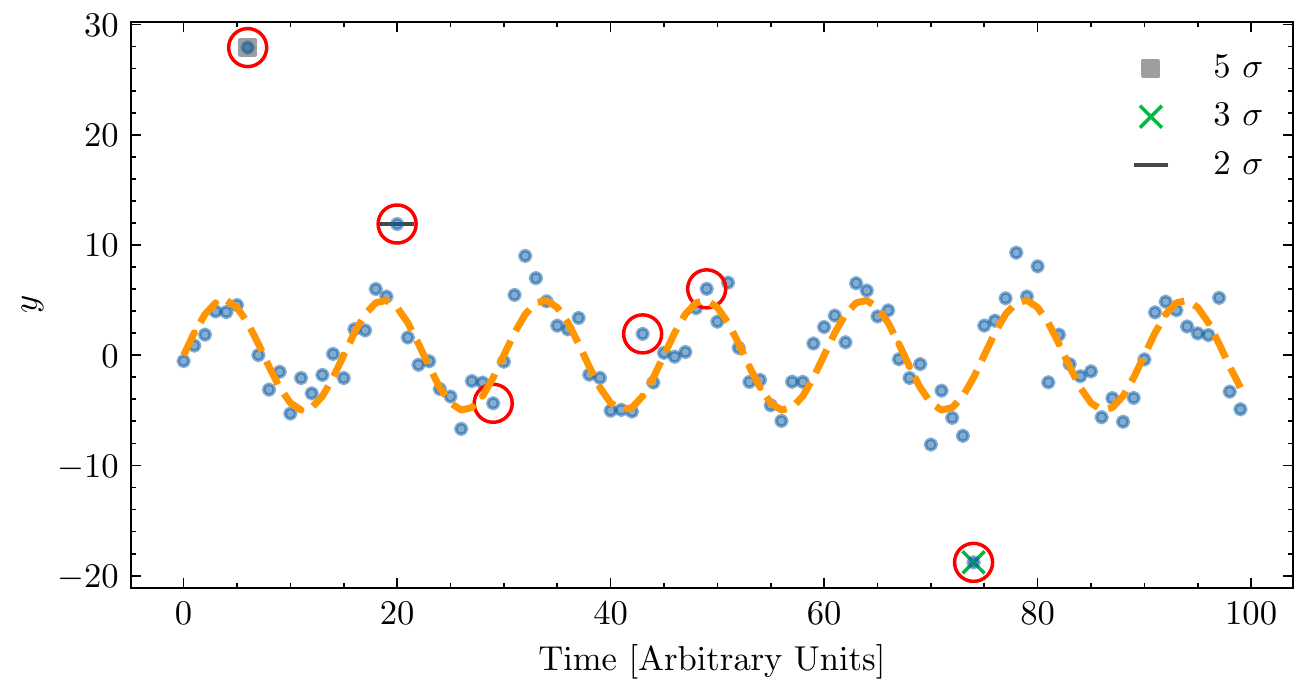}
    \caption{Simulated noisy sine wave data (blue dots) with outliers marked in red, overlayed with the injected sine wave (orange dotted line). Outliers identified by ``sigma clipping'' for $5\sigma, 3\sigma, 2\sigma$ thresholds are marked with a grey square, green cross, and black dashed lines respectively. Even with a generous outlier decision threshold of $2\sigma$, sigma clipping only identifies three outlier points.}
    \label{fig:sigma_clipping}
\end{figure}

Our results are in the bottom panel of \autoref{fig:picture1}, which shows the marginalized outlier probability of each data point, as calculated using techniques from \citet{Vallisneri_2017}. Our results indicate that if we set our outlier probability threshold to be $75\%$ then all four identified outliers are true outliers, and four of the six real outliers are correctly identified. While methods such as sigma clipping are expected to be able to flag outliers $1$, $2$ and $6$ (from left to right), the $4$th outlier could be challenging for that method. We explore this in \autoref{fig:sigma_clipping}, which shows the result of sigma clipping for points more than $2\sigma, 3\sigma$, and $5\sigma$ away, respectively. By definition, points that are $5\sigma$ away are also beyond $3\sigma$ and $2\sigma$ thresholds. Thus, outlier $1$ is identified by $5\sigma$, $3\sigma$, and $2\sigma$ clipping methods. Outlier $6$ is flagged by $3\sigma$ and $2\sigma$ methods, and outlier $2$ is only identified by clipping points $2\sigma$ away. While our Bayesian methods successfully identified four outliers, sigma clipping with a $2\sigma$ threshold could only identify three points. This shows the advantage of our method in that it accounts for other data variations and dependencies, rather than just relying on the measurement uncertainty of the data point. More broadly, one of the key benefits of our Gibbs method is that the marginalized outlier probability not only provides the ability to better understand the severity of potential outliers, but also the freedom of choosing desired outlier thresholds to make informed decisions about possible data excision.

\section{Results}\label{sec:results}

\subsection{Pulsar timing data} \label{sec:pulsar_timing}

Pulsar timing arrays (PTAs) consist of ensembles of well-timed millisecond pulsars in the Milky Way, having the ability to search for low-frequency gravitational waves \citep{1990ApJ...361..300F} from supermassive black-hole binaries, and possible exotic sources such as cosmic strings and dark matter \citep{2019A&ARv..27....5B}. In these searches, PTAs seek unexplained residual contributions to the radio-pulse times of arrival (TOAs) after subtracting a best-fit deterministic timing ephemeris from the data. This timing ephemeris includes such physical dependencies as the pulsar's rotational period (and its evolution), astrometric time delays, radio frequency-dependent delays from interstellar propagation effects, advances due to the relative motion of the pulsar system with respect to the Earth, and (for pulsars in binary systems) orbital kinematic and light propagation effects \citep{2004hpa..book.....L}. The timing residuals are modeled as
\begin{equation}
    \delta\boldsymbol{t} = M\epsilon + F\boldsymbol{a} + U\boldsymbol{j} + \boldsymbol{n},
\end{equation}
where $\delta \boldsymbol{t}$ represents the timing residuals that are decomposed into several components. The term $M\epsilon$ represents linear departures from the best-fit timing ephemeris, where $M$ is the timing ephemeris design matrix that consists of partial derivatives of the TOAs with respect to timing parameters, and $\epsilon$ are linear timing parameter offsets. The term represents $F\boldsymbol{a}$ all low-frequency signals that are modeled with a limited number of Fourier modes, where the matrix $F$ has alternating sine and cosine basis functions at each sampling frequency in the time-series, and $\boldsymbol{a}$ are Fourier coefficients.  
The term $U\boldsymbol{j}$ describes white noise (temporally uncorrelated) that is fully correlated across different radio observing frequencies in near-simultaneous observations. The matrix $U$ maps the number of $\sim 1$~second-duration epochs to the number of TOAs by forming a structure of $N_\mathrm{TOA} \times  N_\mathrm{epoch}$, while $\boldsymbol{j}$ represents the white noise vector. 

The remaining term $\boldsymbol{n}$ describes all other white noise contributions after having accounted for possible systematics. This remaining white noise is characterized by several parameters: there is an EFAC parameter that scales all TOA uncertainties, an additional EQUAD parameter that signifies noise added in quadrature to the scaled uncertainties, and finally ECORR that captures the correlated noise in all TOAs that fall within a given epoch. All of these terms are statistically independent between pulsars, and there are multiple instances of each parameter in a given pulsar that characterizes the noise in each distinct receiver-backend telescope combination. The non-ECORR white noise is then characterized by its covariance between TOAs as
\begin{equation}
    \left\langle n_{i} n_{j}\right\rangle=\sum_{k} N_{i j, k}=\sum_{k} E_{k}^{2}W_{ij} + Q_{k}^{2}\delta_{ij},
\end{equation}
where $E_k$ is EFAC in receiver-backend combination $k$, $Q_k$ is EQUAD, $W = \mathrm{diag}\{\sigma^2\}$ are the squared TOA uncertainties. 
ECORR is modeled separately in an analogous fashion to how low-frequency processes are modeled, which is explained further below.  

Concatenating all data from all pulsars, the PTA likelihood function can be written as
\begin{equation}
    p(\delta \boldsymbol{t} \mid \boldsymbol{\epsilon}, \boldsymbol{a}, \boldsymbol{j}, \phi)=\frac{\exp \left(-\frac{1}{2} \boldsymbol{r}^{T} N^{-1} \boldsymbol{r}\right)}{\sqrt{\operatorname{det}(2 \pi N)}},
\end{equation}
where $\boldsymbol{r}$ denotes our model-dependent  approximation of the white-noise $\boldsymbol{n}$, defined as
\begin{equation}
    \boldsymbol{r}=\delta \boldsymbol{t}-M\boldsymbol{\epsilon}-F \boldsymbol{a}-U \boldsymbol{j},
\end{equation}
and $N=\sum_{k} N_{k}$ is the white noise covariance matrix. The parameter $\phi$ describe any other parameters in addition to the coefficients $\boldsymbol{\epsilon}$, $\mathbf{a}$, $\boldsymbol{j}$. The various design matrices and coefficients are grouped together to write the data structure more succinctly as
\begin{equation}
    T=\left[\begin{array}{lll}
    M & F & U
    \end{array}\right], \quad \mathbf{b}=\left[\begin{array}{c}
    \epsilon \\
    \boldsymbol{a} \\
    \boldsymbol{j}
    \end{array}\right],
\end{equation}
and the coefficients $\boldsymbol{b}$ are further constrained by modeling them as a Gaussian process, with prior
\begin{equation}
    B=\left[\begin{array}{lll}
            \infty & 0 & 0 \\
            0 & \varphi & 0 \\
            0 & 0 & \mathcal{J}
    \end{array}\right]
\end{equation}
where the linear timing-ephemeris perturbations are essentially modeled with an unconstrained uniform prior through an infinite variance Gaussian, while the prior variances on low-frequency noise and ECORR are $\varphi$ and $\mathcal{J}$, respectively. The prior variance on the low-frequency noise corresponds to a (hyper-)parametrized model of the noise's power spectral density (usually a power-law across frequencies). The prior variance on the ECORR is modeled as a parameter that is varied for each distinct receiver-backend combination, exactly as for EFAC and EQUAD. This imbues our PTA model with the following hierarchical structure
\begin{align}
    \delta \boldsymbol{t} \mid \boldsymbol{b} &\sim \mathcal{N}\left(T \boldsymbol{b}, N\right), \nonumber\\
    \boldsymbol{b}  &\sim \mathcal{N}(\boldsymbol{0}, B).
\end{align}

To implement our Gibbs outlier scheme within this model, we rewrite the data structure as
\begin{align}
    \delta \boldsymbol{t} \mid \boldsymbol{b} &\sim \mathcal{N}\left(T \boldsymbol{b}, N_\alpha\right), \nonumber\\
    \boldsymbol{b} &\sim \mathcal{N}(0, B), \nonumber\\
    z_{i} \mid \theta, &\sim \operatorname{Bernoulli}\left(\theta\right), \nonumber\\ 
    \theta &\sim \operatorname{Beta}(km,k(1-m)), \nonumber\\ 
    \alpha_{i} &\sim \operatorname{InverseGamma}(\nu / 2, \nu / 2),
\end{align}
where $N_\alpha = \alpha^{z_{i}}_{i} N$ is introduced as a compact notation for the scaled noise covariance matrix. Notice that the variables used here to characterize the data uncertainties are the same as in \autoref{eq:out_model}. The variable $k$ in the Beta prior on $\theta$ represents the length of the data vector; $m$ is assigned a value of $0.01$, which reflects our confidence that the fraction of outlying observations is small. Given  the relatively large dataset we are working with in the pulsar-timing scenario, and the noise models that we have already implemented, this value of $m$ (and the other choices in the hierarchical outlier model) are appropriate. However, see \citet{Ellis_2019} for further details on how these choices are made in situations with smaller/larger data volume and varying levels of outlier contamination. 

Now we must categorize our model parameters based on whether or not they have an easily-derivable conditional posterior probability distribution that corresponds to standard families of probability distributions. As previously explained, if they do not, then we conduct a short Metropolis-Hastings MCMC chain with other parameters fixed, after which we select the final value as the relevant random draw from the conditional posterior. Building on the existing PTA hierarchical modeling framework, our PTA outlier model parameters have the following conditional posterior distributions:
\begin{widetext} 
\begin{align} \label{eq:PTA_update}
    \pi_{1}(N\mid \boldsymbol{\eta},\delta \boldsymbol{t}) &\sim \mathrm{MetropolisHastings} \nonumber\\
    \pi_{2}(B\mid \boldsymbol{\eta},\delta \boldsymbol{t}) &\sim \mathrm{MetropolisHastings} \nonumber\\
    \pi_{3}(\boldsymbol{b}\mid \boldsymbol{\eta},\delta \boldsymbol{t}) &\sim \mathcal{N}\left( \left(T^{T}N_\alpha^{-1}T+B^{-1}\right)^{-1} T^T N_\alpha^{-1}\delta\boldsymbol{t},\,\,\left(T^{T}N_\alpha^{-1}T+B^{-1}\right)^{-1}\right) \nonumber\\
    \pi_{4}(\theta \mid \boldsymbol{\eta},\delta \boldsymbol{t}) &\sim \operatorname{Beta}\left(km+\sum_{i=1}^{n} z_{i}, k(1-m)+n-\sum_{i=1}^{n} z_{i}\right) \nonumber\\
    \pi_{5}(z_{i} \mid \boldsymbol{\eta},\delta \boldsymbol{t})&\sim \operatorname{Binomial}\left(\frac{\theta_{i}\times \mathcal{N}(T\boldsymbol{b}, N_\alpha)}{\theta_{i}\times \mathcal{N}(T\boldsymbol{b}, N_\alpha) + (1-\theta_{i}) \times \mathcal{N}(T\boldsymbol{b}, N)}\right) \nonumber\\
    \pi_{6}(\alpha_{i} \mid \boldsymbol{\eta},\delta \boldsymbol{t}) &\sim \operatorname{InverseGamma}\left(\frac{1}{2}(\nu + z_{i}), \frac{1}{2}\left(\nu + z_{i}(\delta \boldsymbol{t}-\mathrm{T}\boldsymbol{b})^{T}N^{-1}(\delta \boldsymbol{t}-\mathrm{T}\boldsymbol{b})\right)\right)
\end{align}
\end{widetext}

\subsubsection{Simulated data}

\begin{figure} \label{figure2}
	\includegraphics[width=\columnwidth]{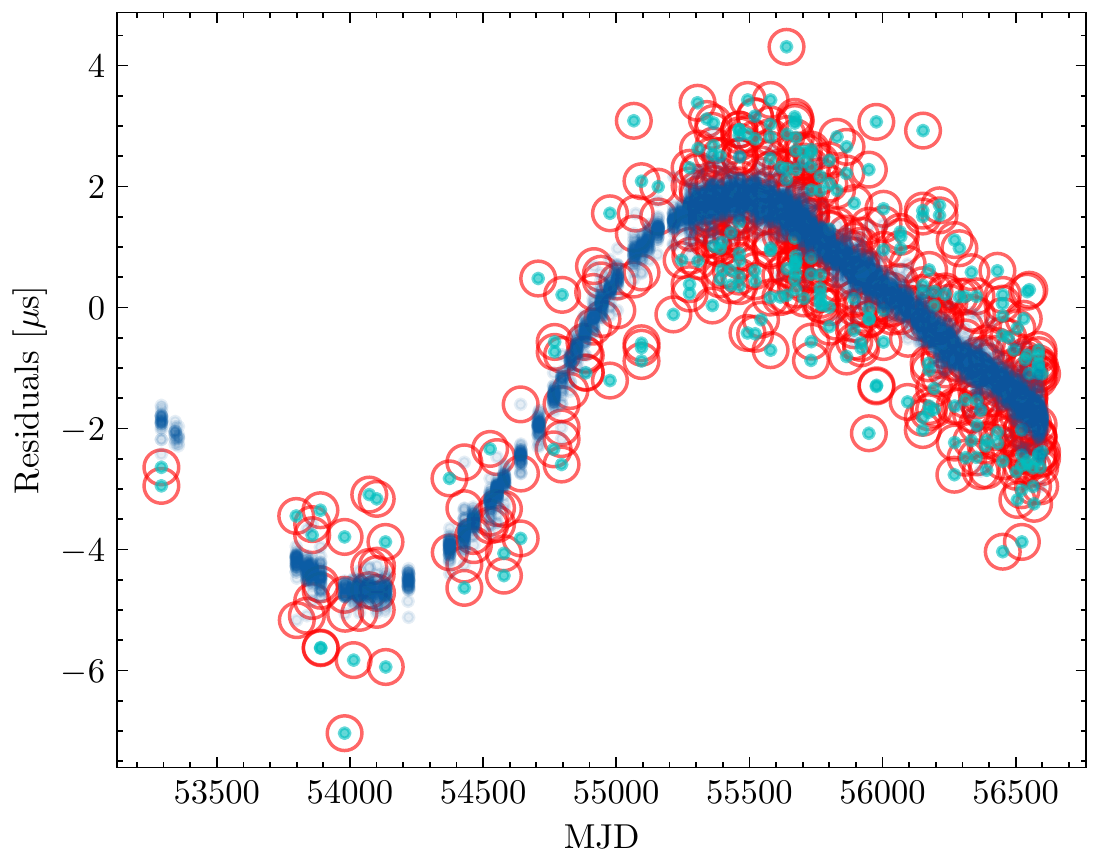}
    \caption{Simulated J$1909$$-$$3744$ pulsar timing data, with inlier (dark blue) and random outliers (cyan) consist of $5\%$ of the total population. Outliers identified by our sampling procedure are circled in red.}
    \label{fig:picture2}
\end{figure}

We test this outlier scheme on $\sim9$~years of simulated timing data from pulsar J$1909$$-$$3744$\footnote{\href{https://github.com/jellis18/gibbs_student_t/blob/master/simulate_data.py}{https://github.com/jellis18/gibbs\_student\_t/blob/master/simulate\_data.py}} using the {\tt Libstempo}\footnote{\href{https://github.com/vallis/libstempo.git}{https://github.com/vallis/libstempo.git}} software package. We assume that outliers constitute $5\%$ of the total data, and randomly assign datapoints to be outliers according to this. The inlier data model consists of the pulsar timing ephemeris, intrinsic red noise, TOA measurement uncertainties, and EQUAD parameters for each distinct receiver-backend combination. While no EQUAD is injected, it is modeled with a prior distribution that is uniform in log space, i.e., $p(\log_{10}(\mathrm{EQUAD} / \mathrm{sec}))= U[-10,-4]$. Intrinsic pulsar red noise is modeled with a power-law power spectral density, such that $P(f) = A^2 f^{-\gamma} / 12 \pi^{2}$, with priors $p(\log_{10}A)=U[-18,-11]$ and $p(\gamma) = U[0,7]$, and injected values $A = 10^{-14}$ and $\gamma = 13/3$. 
The simulated dataset is shown in \autoref{fig:picture2}, with dark blue indicating inlier residual datapoints, and cyan indicating outlier datapoints.

The Gibbs outlier scheme outlined in the previous section is then applied to this dataset, where for each step in the Gibbs update loop we proceed through all six steps in \autoref{eq:PTA_update} sequentially. For those updates that don't have conditional posteriors corresponding to standard families of distributions, we carry out short MCMC runs for $20$ Metropolis-Hastings steps, taking the final parameter position as the updated value. In total, the Gibbs loop is iterated through $10^4$ times to provide a set of random draws from the joint posterior distribution of all inlier and outlier model parameters. 

\begin{figure}
	\includegraphics[width=\columnwidth]{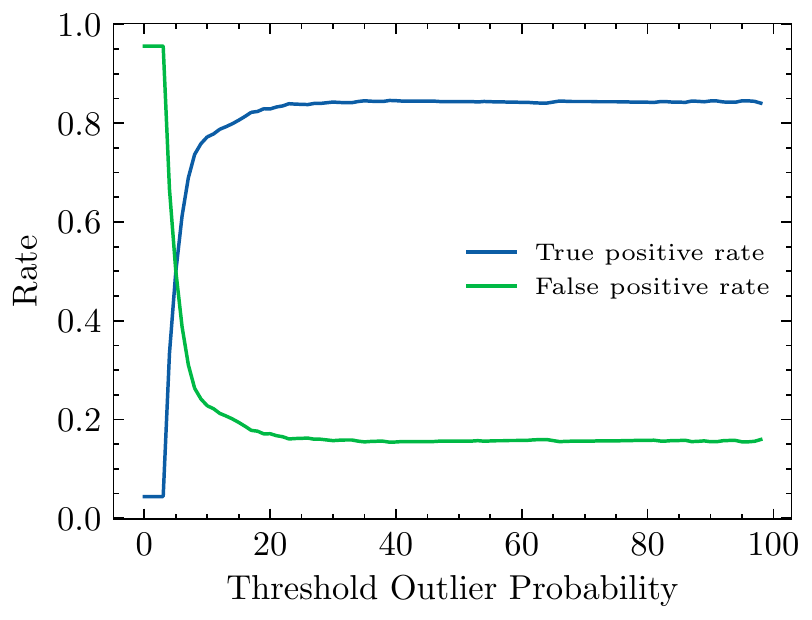}
    \caption{True positive rate calculated as ratio of correctly identified outliers to total outliers. False positive rate calculated as ratio of falsely identified outliers to total outliers.}
    \label{fig:picture3}
\end{figure}

In \autoref{fig:picture2} we circle those points in red that are identified as having a marginalized outlier probability $>95\%$, where this is again calculated as in \citet{Vallisneri_2017}. We see that most true outliers (cyan) are correctly identified with this threshold. However, for a more robust and systematic analysis, we calculate the true positive rate and the false positive rate of outlier identification as a function of the marginalized outlier probability threshold. Our results are shown in \autoref{fig:picture3}, where they indicate that the true positive rate (the percentage of correctly identified outliers out of all outliers) is $\sim4$ times larger than the false positive rate (falsely identified outliers, out of all outliers), meaning we identified four times more true outliers than false outliers. This gives confidence that our model is able to achieve robust outlier mitigation. 

\subsubsection{NANOGrav $9$-year Dataset: $\mathrm{J}1909$$-$$3744$}

\begin{figure*}
	\includegraphics[width=0.8\textwidth]{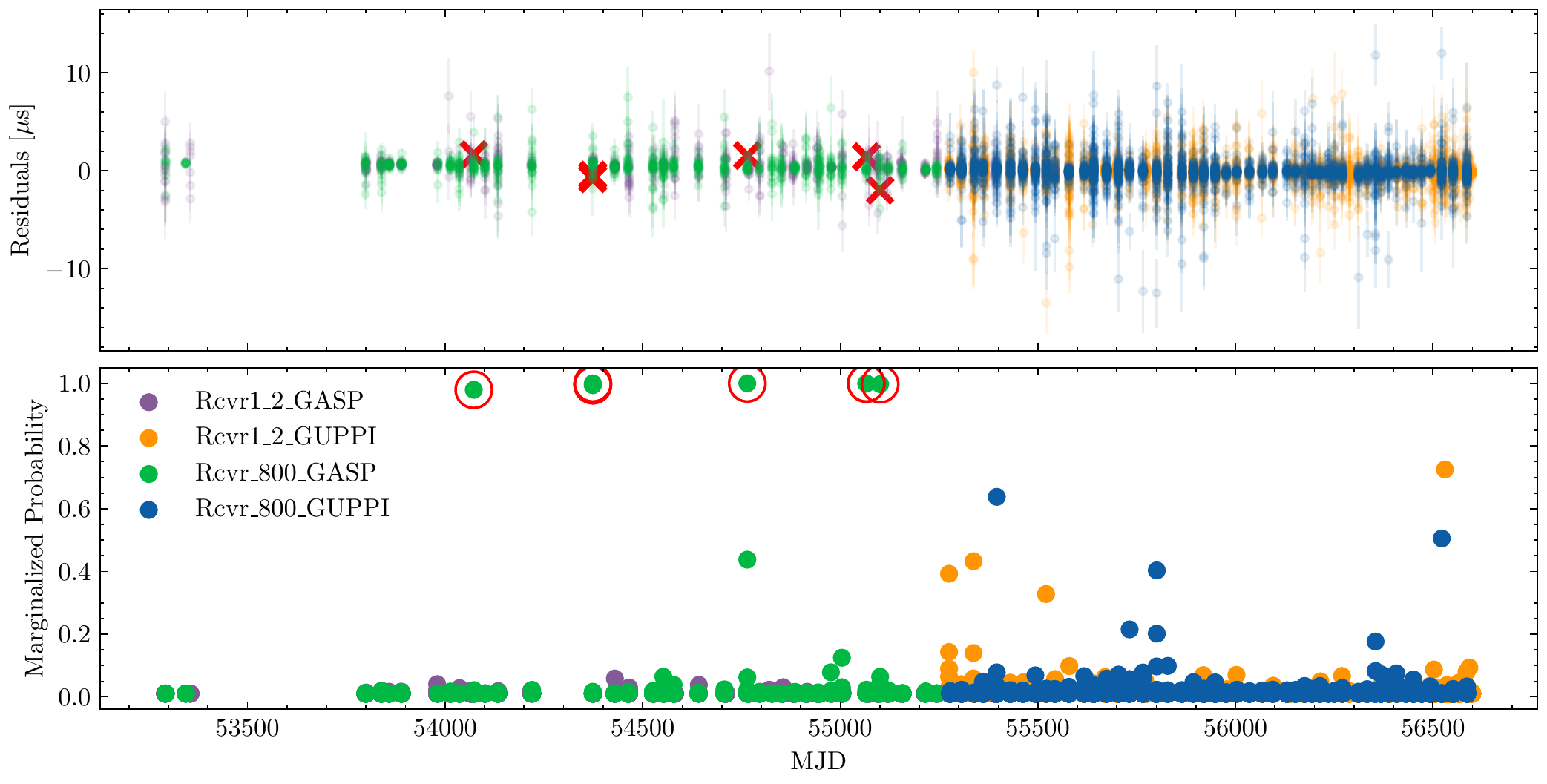}
    \caption{\textit{Top:} timing residuals for pulsar J$1909$$-$$3744$ from the NANOGrav $9$-year Dataset are colored coded by different combinations of receiver and timing backend. Data identified by our Gibbs modeling scheme as having a marginalized outlier probability in excess of $95\%$ are indicated by red cross. \textit{Bottom:} Marginalized outlier probability of each datapoint, color coded as in the top panel. Points with marginalized outlier probability in excess of $>95\%$ are circled in red, with all seeming to belong to the same combination of receiver and timing backend ($800$-GASP).}
    \label{fig:picture4}
\end{figure*}

\begin{figure*}
	\includegraphics[width=0.8\textwidth]{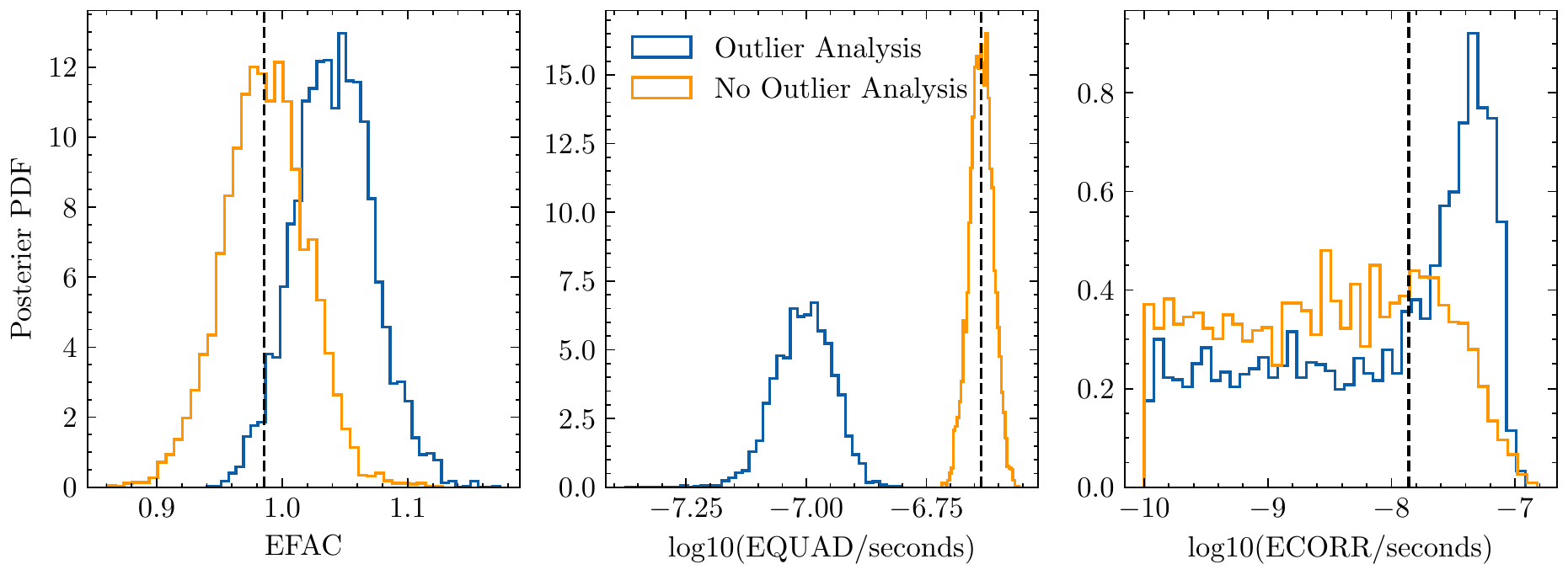}
    \caption{Each panel shows the marginalized $1$D posterior probability distribution of white noise parameters for $800$~Mhz data in the GASP timing backend. The data are for pulsar J$1909$$-$$3744$ from the NANOGrav $9$-year Dataset. Blue distributions derive from our Gibbs outlier analysis, while orange distributions are from a standard analysis where outliers are untreated. Dashed vertical lines correspond to values reported in \citet{2015_NANOGrav9yr}.}
    \label{fig:picture5}
\end{figure*}

We now apply our techniques to real pulsar observations from the NANOGrav $9$-year Dataset \citep{2015_NANOGrav9yr}. The same pulsar is analyzed as was simulated previously, i.e., J$1909$$-$$3744$, with our results contrasted aginst a standard analysis of the data that does not account for outliers. The white noise parameters in the data model are assigned the following priors: $p(\mathrm{EFAC}) = U[0.1,10]$, $p(\log_{10}(\mathrm{EQUAD} / \mathrm{sec}))= U[-10,-4]$, and $p(\log_{10}(\mathrm{ECORR} / \mathrm{sec}))= U[-10,-4]$, where these are again modeled for each distinct receiver-backend combination. Intrinsic red noise is also modeled with the same priors as in our simulated data tests. Likewise, our Gibbs updating scheme is exactly as previously described.  

\autoref{fig:picture4} shows the results of our outlier analysis, where the top panel displays timing residuals that are color coded by each receiver-backend combination, and points that are identified as having $>95\%$ marginalized outlier probability are labeled with a red cross. The lower panel explicitly shows the marginalized outlier probability for each data point, again color coded by receiver-backend combinations, and the same points that exceed the $95\%$ threshold indicated by a red circle. Due to the proximity of the identified outliers, \autoref{fig:picture4} contains seven total outliers, some with overlapping crosses/circles. We see that all data points that exceed this threshold belong to the combination of data observed at $800$~MHz with the GASP backend. If unaccounted, these data could act as potential sources of bias in the statistical inference of parameters in our model. 

One of the benefits of our Bayesian outlier mitigation scheme is that we are able to re-calibrate the statistical inference of model parameters that would otherwise be contaminated by the untreated presence of outliers. We demonstrate this in \autoref{fig:picture5}, which shows the marginalized $1$D posterior probability distributions of EFAC, EQUAD, and ECORR parameters for the $800$-GASP receiver-backend data. In each panel, the results of our analysis that include outlier mitigation (blue) are contrasted against a standard analysis that does not account for outliers (orange), and whose joint posterior was sampled using parallel-tempering Metropolis-Hastings MCMC techniques. We observe that with outlier mitigation leads to a significant decrease in measured EQUAD---more than $50\%$, judging by maximum-a-posterior values---which is additional white noise that adds in quadrature to the usual TOA uncertainties. The fact that less white noise is needed to explain the data when we account for outliers in $800$-GASP makes sense; the untreated presence of outliers has the effect of inflating the EQUAD parameter to explain their presence as part of the model. A reduced EQUAD value signals that we have successfully diagnosed and mitigated the influence of outlier data, and for subsequent analysis we are able to have more accurate inference on other parameters of interest. However note that no data has been explicitly excised, only probabilistically isolated away from the inlier model inference. 

This analysis has broader benefits than just lowering the inferred level of noise in the radio receiver-backends suffering from outlier contamination. By localizing the origin of this contamination to a known piece of hardware or post-processing software, further investigations could be made into the epochs when these outliers occurred, as they may correspond to known instrumental issues. Furthermore, isolating outlier data points---that would otherwise inflate the white noise level---could help to avoid spurious contamination in the search for continuous GW signals from individual SMBHBs.

\begin{figure*}
	\includegraphics[width=0.8\textwidth]{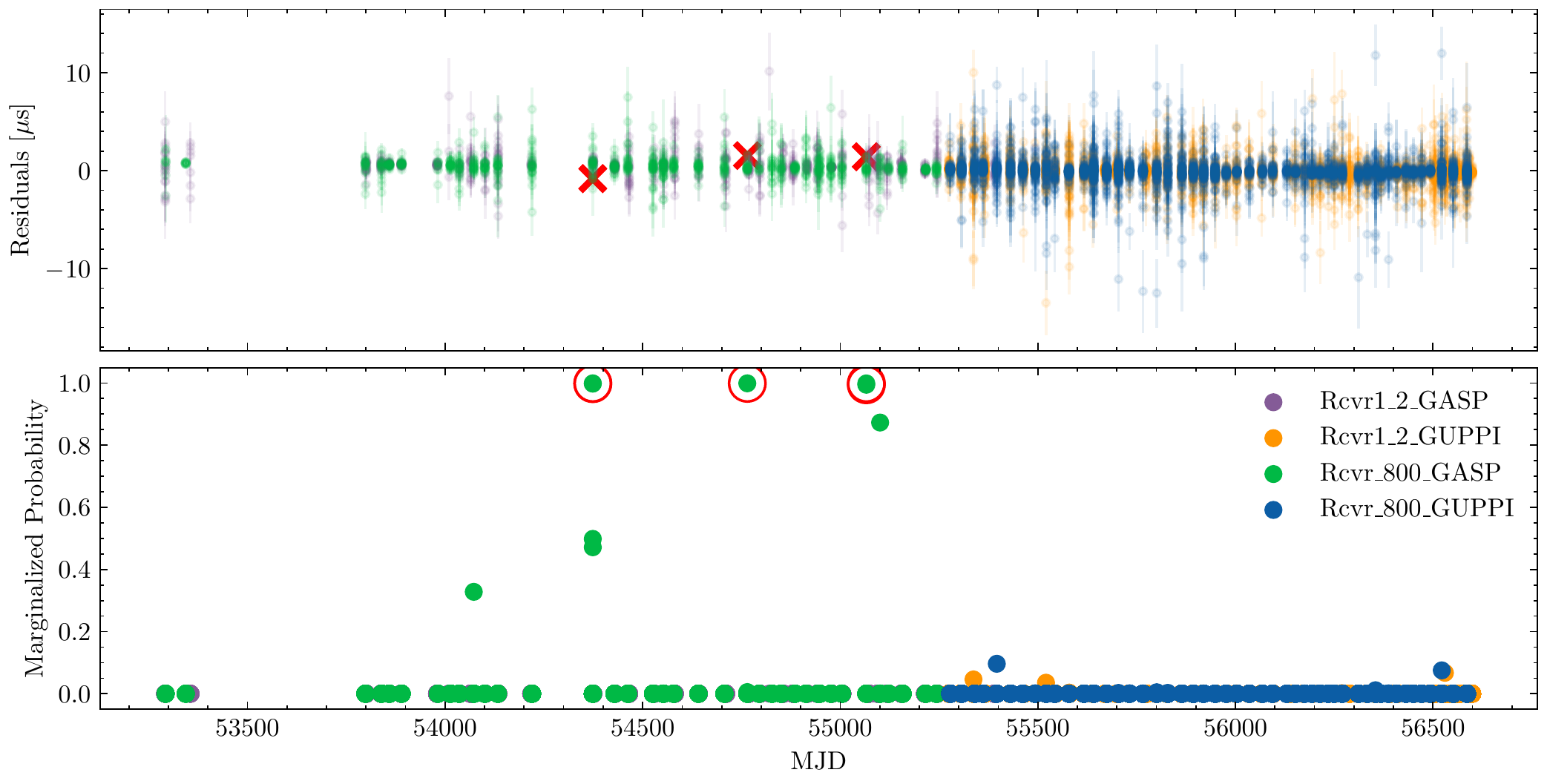}
    \caption{\textit{Top:} timing residuals for pulsar J$1909$$-$$3744$ from the NANOGrav $9$-year Dataset are analyzed with the \textit{vvh17} modeling scheme \citep{Vallisneri_2017}, and those having a marginalized outlier probability in excess of $95\%$ are indicated by red cross. \textit{Bottom:} Marginalized outlier probability of each datapoint, color coded as in the top panel. Points with marginalized outlier probability in excess of $>95\%$ are circled in red, with all seeming to belong to the same combination of receiver and timing backend ($800$-GASP).}
    \label{fig:vvh17toa}
\end{figure*}

Finally, we compare our method to another Bayesian outlier-robust technique for pulsar-timing datasets, described in \citet{Vallisneri_2017} and hereafter referred to as \textit{vvh17}. In that model, outliers are posited to be distributed uniformly across a given pulsar's rotational period. The posterior sampling of the full hierarchical model is tackled using Hamiltonian Monte Carlo in that paper, but it is possible to embed their same outlier modeling assumptions within our Gibbs-sampling framework. In other words, we compared the performance of our outlier model to that of \textit{vvh17} (albeit implemented with a Gibbs sampler instead of Hamiltonian Monte Carlo). The results of this are shown in \autoref{fig:vvh17toa}, where the top panel shows timing residuals, color-coded by each receiver-backend combination, while the bottom panel displays points that are tagged for having $>95\%$ marginalized outlier probability. Due to the proximity of the identified outliers, \autoref{fig:vvh17toa} contains five total outliers, some with overlapping crosses/circles. We cross-examined outliers identified by our method with \textit{vvh17}, and found that the five outliers tagged by \textit{vvh17} are a subset of the seven that our method deduced. This demonstrates that our proposed method delivers consistent results with other Bayesian outlier methods in a pulsar-timing context, and that our method does a more thorough job of outlier isolation. 

\begin{figure*}
	\includegraphics[width=0.8\textwidth]{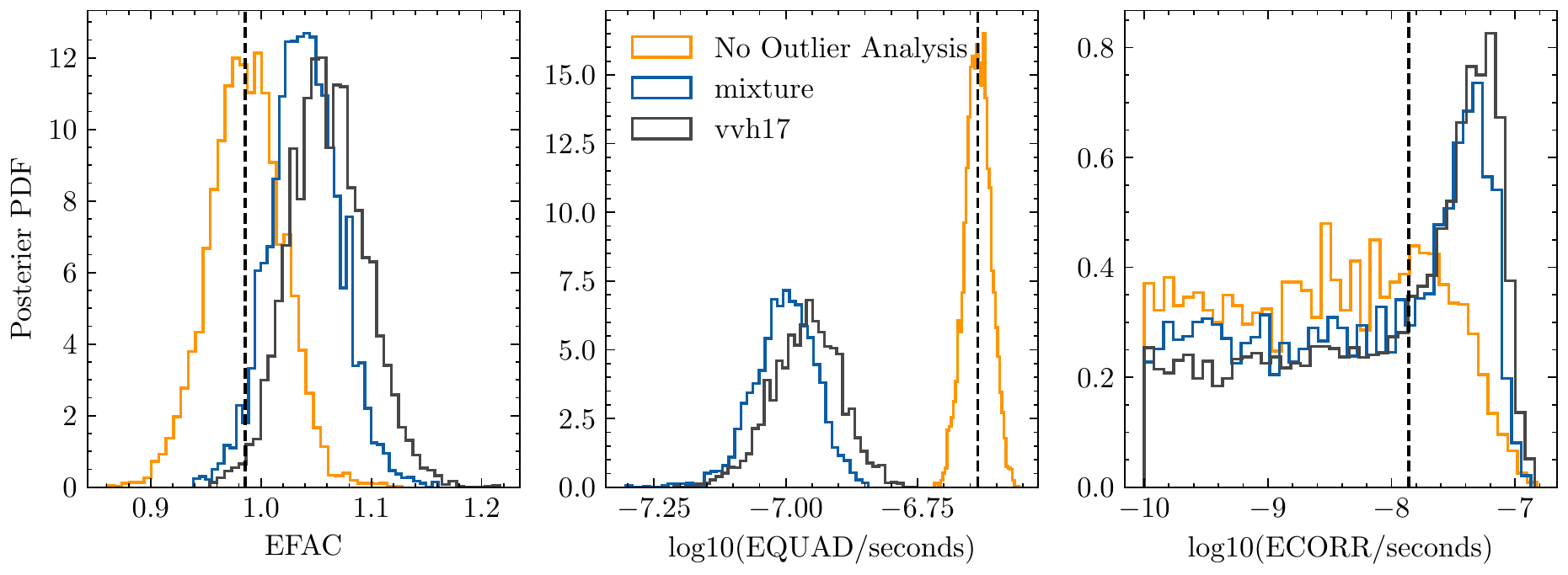}
    \caption{Each panel shows the marginalized $1$D posterior probability distribution of white noise parameters for $800$~Mhz data in the GASP timing backend. The data are for pulsar J$1909$$-$$3744$ from the NANOGrav $9$-year Dataset. Blue distributions derive from our Gibbs outlier analysis, black distributions are results from \textit{vvh17} method \citep{Vallisneri_2017}, while orange distributions are from a standard analysis where outliers are untreated. Dashed vertical lines correspond to values reported in \citet{2015_NANOGrav9yr}.}
    \label{fig:vvh17parameters}
\end{figure*}

In addition, the bottom panel of \autoref{fig:vvh17toa} shows that \textit{vvh17} also finds that the $800$-GASP receiver-backend combination is a source of outliers, adding further validation to the results of our method. We again compare the marginalized $1$D posterior probability distributions of EFAC, EQUAD, and ECORR parameters for the $800$-GASP receiver-backend data in \autoref{fig:vvh17parameters}. In each panel, we present results from our proposed method (blue, labeled as \textit{mixture}), the \textit{vvh17} method (black), and a standard analysis that does not include outlier mitigation methods (orange). We first notice the increase in EFAC parameter from the \textit{vvh17} method, which shows our proposed method can remove more noise. Similarly, we observe an increase in EQAUD from \textit{vvh17}, though both methods are able to have a significantly reduced EQUAD value compared to the analysis that does not have any outlier mitigation. Overall, it shows that both techniques effectively combat the influence of outlier contamination and re-calibrate the model parameters. Nevertheless, our proposed method is a more conservative outlier modeling approach, identifying several more outliers, and having a greater noise-reduction influence.

\subsection{Time-domain photometric light-curves} \label{sec:photo_data}

\subsubsection{Photometric data}

\begin{figure*}
	\includegraphics[width=0.8\textwidth]{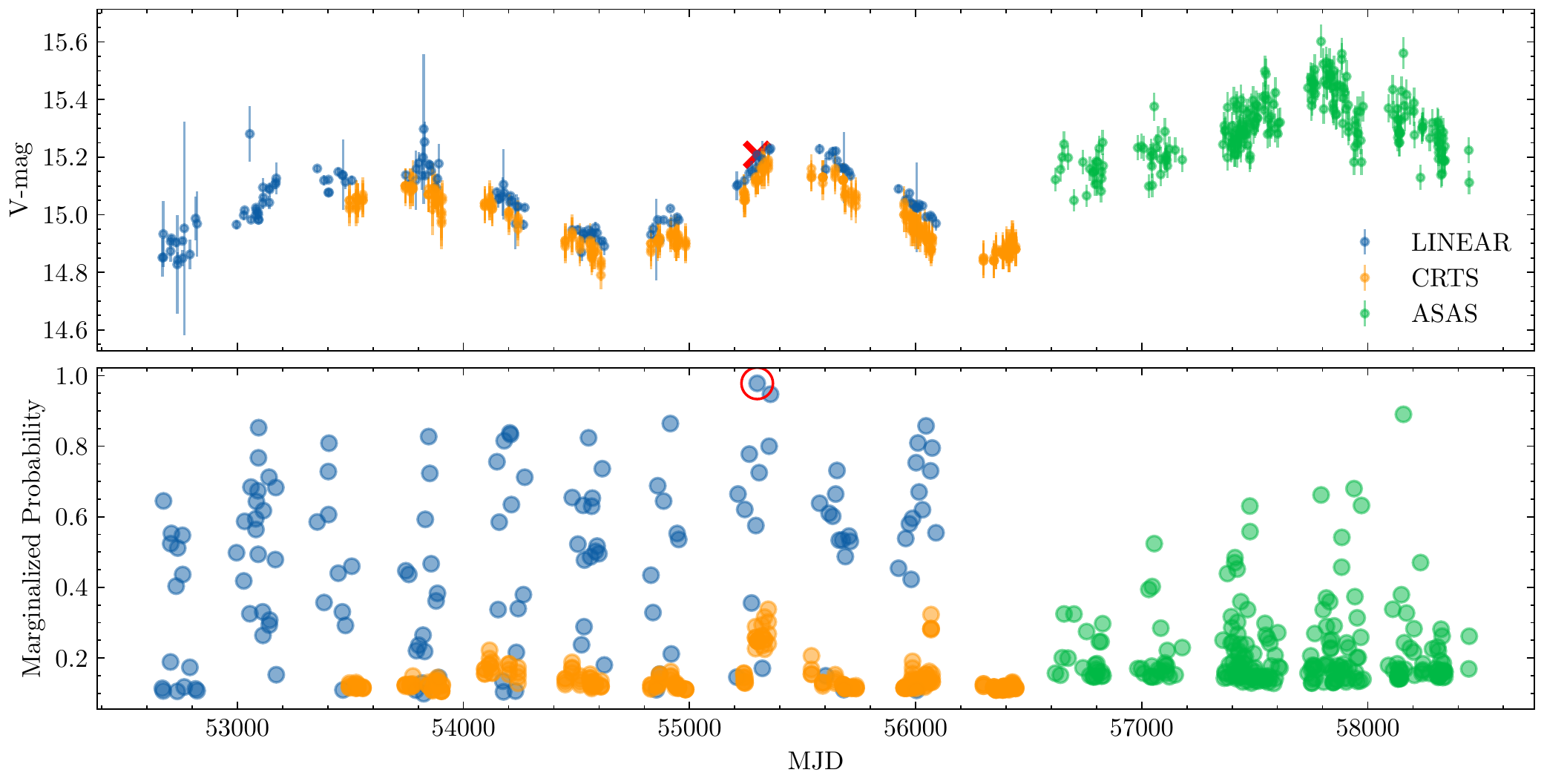}
    \caption{\textit{Top:} Lightcurve magnitudes for AGN candidate PG $1302$-$102$ are color-coded by different survey. Data identified by our Gibbs modeling scheme as having a marginalized outlier probability in excess of $95\%$ are indicated by red cross. \textit{Bottom:} Marginalized outlier probability of each datapoint, color coded as in the top panel. Points with marginalized outlier probability in excess of $>95\%$ are circled in red.}
    \label{fig:picture6}
\end{figure*}

\begin{figure*}
	\includegraphics[width=0.8\textwidth]{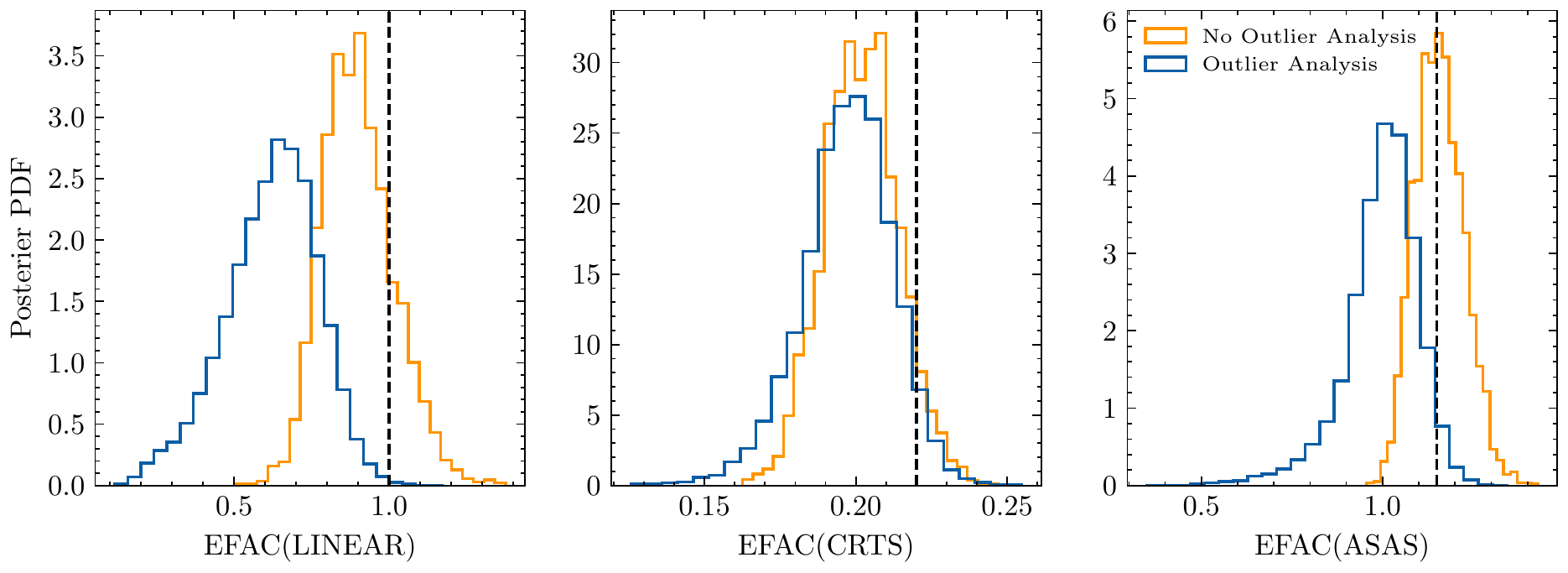}
    \caption{Each panel shows the marginalized $1$D posterior probability distribution of EFAC uncertainty-scaling parameters for PG$1302$$-$$102$ from the combination of CRTS, ASAS-SN, and LINEAR survey data. Blue distributions derive from our Gibbs outlier analysis, while orange distributions are from a standard MCMC analysis where potential outliers are untreated. Dashed vertical lines correspond to values reported in \citet{Zhu_2020}.}
    \label{fig:picture8}
\end{figure*}

Electromagnetic (EM) information offers another pathway to observing closely-separated SMBHB systems. While direct imaging of dual radio cores has not been successful for any sub-parsec separated binaries, time-domain quasar photometry offers an indirect way to infer the presence of a binary through periodic variations that may be tied to the binary's orbital dynamics. 

We write the data model for periodicity searches in quasar time-domain optical magnitude measurements as
\begin{equation}
    \boldsymbol{d} = \boldsymbol{n} + \boldsymbol{m} + \boldsymbol{s},
\end{equation}
where $\boldsymbol{d}$ is the light curve data (typically magnitude measurements in a certain band); $\boldsymbol{n}$ represents the noise vector, which includes measurement uncertainties and additional intrinsic stochastic quasar variability; $\boldsymbol{m}$ describes the mean magnitude with offsets due to host galaxy light contamination; and $\mathbf{s}(t) = A \sin \left(2 \pi f_{0} t+\phi\right)$ describes a periodic signal. We follow the likelihood structuring of \cite{Zhu_2020}, such that:
\begin{equation}
    p(\mathbf{d} \mid \boldsymbol{\vartheta}_{n}, \boldsymbol{\vartheta}_{s}, \boldsymbol{m}) = 
    \frac{\exp \left[-\frac{1}{2}(\mathbf{d}-\mathbf{m}-\mathbf{s})^{T} C^{-1} (\mathbf{d}-\mathbf{m}-\mathbf{s})\right]}{\sqrt{\mathrm{det}(2 \pi C})},
\end{equation}
where $\boldsymbol{\vartheta}_{n}$ and $\boldsymbol{\vartheta}_{s}$ are noise and signal parameters respectively, and $C$ represents the noise covariance matrix
\begin{equation}
    C_{i j} = \left\langle n_{i} n_{j}\right\rangle = E^2 \sigma^2_{i} \delta_{i j} + \frac{1}{2} \hat{\sigma}^{2} \tau_{0} \exp \left[-\left(\frac{\tau_{i j}}{\tau_{0}}\right)^{\gamma}\right],
\end{equation}
where $\sigma_{i}$ is the measurement uncertainty for the $i$th observation, E is a scaling factor used to quantify the over/under-estimation of measurement uncertainties (similar to the pulsar-timing EFAC parameter), and $\delta_{i j}$ is the Kronecker delta function. These three parameters form a diagonal matrix that represents white noise in the data covariance. We model an additional source of noise that represents intrinsic quasar photometric variability (or so called stochastic ``red noise'') as a stretched exponential process (or a Kohlrausch-Williams-Watts process) with the following parameters: $\hat{\sigma}^{2}$ is the intrinsic variance between observations on short timescales; $\tau_{0}$ represents the noise damping timescale; $\tau_{i j} \equiv\left|t_{i}-t_{j}\right|$ is the absolute time difference between observations; and $\gamma$ adds  additional control on the correlation lengthscale of the process. We note that $\gamma=1$ corresponds to the usual quasar variability model of a Damped Random Walk (DRW) \citep{2009ApJ...698..895K}, $\gamma=0$ is white noise, and $\gamma=2$ is a Gaussian function. This stretched exponential process gives flexibility to our red noise modeling, allowing it to deviate from the fiducial DRW model for AGN.

To incorporate our Gibbs-sampling outlier mitigate scheme, we construct a slightly modified version of the Bayesian hierarchical photometric data model as
\begin{align}
    d \mid m, \vartheta_{n}, \vartheta_{s} &\sim \mathcal{N}\left(m+s, C_{\alpha} \right), \nonumber\\
    m, \vartheta_{n}, \vartheta_{s}  &\sim \operatorname{Uniform}, \nonumber\\
    z_i \mid \theta &\sim \operatorname{Bernoulli}\left(\theta\right), \nonumber\\ 
    \theta &\sim \operatorname{Beta}(km,k(1-m)), \nonumber\\ 
    \alpha_i &\sim \operatorname{InverseGamma}(\nu / 2, \nu / 2),
\end{align}
where $C_\alpha = \alpha^{z_{i}}_{i} C$ is (like in the PTA case) a compact notation for the noise covariance matrix that is scaled by an outlier-tag dependent additional factor. During Gibbs sampling of the model parameter space, we update each category of parameter as follows:
\begin{widetext} 
\begin{align} 
    \pi_{1}(m\mid \boldsymbol{\eta},\boldsymbol{d}) &\sim \mathrm{MetropolisHastings}, \nonumber\\
    \pi_{2}(\vartheta_{n}\mid \boldsymbol{\eta},\boldsymbol{d}) &\sim \mathrm{MetropolisHastings}, \nonumber\\
    \pi_{3}(\vartheta_{s}\mid \boldsymbol{\eta},\boldsymbol{d}) &\sim \mathrm{MetropolisHastings}, \nonumber\\
    \pi_{4}(\theta \mid \boldsymbol{\eta},d) &\sim \operatorname{Beta}\left(km+\sum_{i=1}^{n} z_{i}, k(1-m)+n-\sum_{i=1}^{n} z_{i}\right), \nonumber\\
    \pi_{5}(z_{i} \mid \boldsymbol{\eta},d) &\sim \operatorname{Binomial}\left(\frac{\theta_{i} \times \mathcal{N}(m+s,C_{\alpha})}{\theta_{i} \times \mathcal{N}(m+s, C_{\alpha}) + (1-\theta_{i}) \times  \mathcal{N}(m+s, C)}\right),
    \nonumber\\
    \pi_{6}(\alpha_{i} \mid \boldsymbol{\eta},d) &\sim \operatorname{InverseGamma}\left(\frac{1}{2}(\nu + z_{i}), \frac{1}{2}(\nu + z_{i} (d-m-s)^{T}C^{-1}(d-m-s)\right).
\end{align}
\end{widetext}

\subsubsection{SMBHB candidate: $\mathrm{PG}1302$$-$$102$}

We choose one of the most prominent binary candidates among periodically-variable quasars: PG$1302$$-$$102$. This $z=0.2784$ quasar with median $V$-band magnitude of $15.0$ was the most significant periodic candidate identified by \citet{Graham_2015a} among the $111$ found in the $243,500$-quasar Catalina Real-time Transient Survey (CRTS) dataset\footnote{\href{http://crts.caltech.edu/}{http://crts.caltech.edu/}} \citep{2009ApJ...696..870D}. With a $1884$-day period, this system has been interpreted as a binary with Doppler-boosted emission from the secondary black-hole's mini-disk \citep{2015Natur.525..351D,Graham_2015b,2015MNRAS.454L..21C}. However, there is ongoing discussion over whether a binary model is truly favored over just intrinsic stochastic variability that appears periodic by happenstance over a short baseline \citep[e.g.,][]{2016MNRAS.461.3145V,2020MNRAS.496.1683X,2018ApJ...859L..12L}. Regardless of its true nature, we employ this binary candidate as an example to showcase our outlier mitigation approach embedded within the photometric data likelihood.

We consider three sources of photometric measurements for this object: $(i)$ the CRTS data set contains 290 photometric measurements taken between 6 May 2005 and 30 May 2013; $(ii)$ the All-Sky Automated Survey for Supernovae (ASAS-SN) \citep{2019MNRAS.486.1907J} data set contains 232 measurements between 23 November 2013 and 27 November 2018; and finally $(iii)$ the Lincoln Near-Earth Asteroid Research (LINEAR) \citep{2011AJ....142..190S} data set includes 626 measurements taken between 23 January 2003 and 12 June 2012. We use the combination of these measurements from \citet{Zhu_2020}, synthesizing a dataset that spans over 15.84 years with median measurement uncertainties of 0.06, 0.05, and 0.01 mag respectively. \citet{Zhu_2020} further bins the LINEAR data with an interval of $1$ day, thereby preserving stochastic time-correlated variations over timescales greater than this bin size. 
The combined data set that we use is shown in the top panel of \autoref{fig:picture6} as $V$-band magnitudes.

In our analysis of this binary candidate, we consider a fiducial model where no outlier mitigation is performed and the joint parameter distribution is sampled with parallel-tempered MCMC techniques. We also apply our new Gibbs-sampled outlier modeling to contrast the results. We place priors on parameters as shown in \autoref{tab:priors}. We found that we needed to increase the number of Gibbs iterations to $5\times 10^4$ in order to reach convergence. However, we continue to perform 20 MCMC iterations in the Gibbs steps for which Metropolis-Hastings sampling of conditional posteriors is needed.
 
The bottom panel of \autoref{fig:picture6} shows our results in the form of the marginalized outlier probability of each data point, indicating that the bulk of potential outliers at each possible threshold cut derive from the LINEAR dataset. One LINEAR datapoint has a marginalized outlier probability in excess of $95\%$, although there are $13$ in excess of $80\%$. Of these $13$, only one is from ASAS-SN and the rest are from LINEAR, indicating that LINEAR has the lowest data quality and CRTS has the highest data quality. To further investigate whether the LINEAR survey might be suffering from lower data quality and a greater number of statistical outliers, in \autoref{fig:picture8} we compare the recovered marginalized posterior distributions of the EFAC parameters in the case where no outliers are modeled (orange histograms; parallel-tempered MCMC) and where outliers are modeled (blue histograms; Gibbs-sampled outlier model). We observe a significant decrease in the LINEAR EFAC parameter (i.e., the scaling parameter of the data measurement uncertainty) when outliers are modeled. This indicates that the binning of LINEAR data -- and the corresponding recalculated uncertainty on the binned data -- is affected by the presence of outliers, causing the EFAC parameter to bloat in order to be consistent with the data fluctuations. This interpretation aligns with the relatively high outlier probabilities for the LINEAR data. Likewise, in the ASAS-SN data we notice a decrease in the EFAC parameter when outliers are tagged and marginalized over. However, the CRTS survey data has some of the lowest data outlier probabilities, which is reflected in the fact that the EFAC parameter posterior does not differ significantly when outliers are modeled. 

Our analysis demonstrates the usefulness of outlier modeling as a diagnostic check on data quality, and to point to possible sources of contamination that may lead to biased inference. More specifically, the presence of untamed outliers in such photometric AGN datasets could have a deleterious impact on the search for periodicity, which is one way that close-separation SMBHB candidates are found. By isolating such outliers, we can improve data quality and reduce instances of false-positive periodicity discovery.

\section{Conclusions}\label{sec:conclusions}

The inference of model parameters from data can be biased by the presence of contaminants, or any source of data that departs significantly from assumptions in our data model. In time-domain applications this can be a systemic process that persists throughout the data collection (e.g., not accounting for time-correlated noise), or transient fluctuations that impact the quality of only a few data points (e.g., radio frequency interference in pulsar timing). In this paper, we studied several time-domain applications where outliers may be present and lead to biased inference. Our core interest has been multi-messenger searches for supermassive black-hole binaries, where outliers and poor data quality in AGN photometry or pulsar-timing datasets can impact the significance of EM and GW searches for these systems. 

We tested a hierarchical Bayesian modeling scheme that considers measured data to be described by a mixture of an inlier model and a (potentially quite different) outlier model. In its simplest form this can be written as a Gaussian mixture model with binary tags for each data point that indicate its outlier status at a given stage in the probabilistic modeling scheme. This outlier model (first introduced in \citet{Ellis_2019}) is most efficiently implemented through Gibbs sampling of the joint parameter posterior distribution, where parameters are divided into groups for which the conditional posterior distribution can be written in the form of a standard distribution family. As such, no rejection-based sampling is needed except in the circumstances where a parameter's conditional distribution is non-standard. In these cases, a short Metropolis-Hastings Markov Chain Monte Carlo chain takes the role of generating a quasi-independent random draw from the conditional distribution. By cycling through parameters and parameter blocks over many iterations in this Gibbs scheme, the joint posterior distribution of our large outlier model can be sampled. Given that each data point is given its own outlier tag, the total number of parameters exceeds the number of data points. However, this is all well behaved through appropriate application of Bayesian priors to control the parameter space.

To demonstrate the versatility of this outlier modeling approach for time-domain observations, we deployed it on two kinds of multi-messenger data streams: pulsar-timing observations that are used for nanohertz-frequency GW searches, and optical magnitude AGN photometry that is used in periodicity searches for black-hole binary systems. For pulsar-timing data, we tested the efficacy of our outlier approach on simulated data of pulsar J$1909$$-$$3744$, finding good performance where the rate of true outlier identification was four times that of erroneous outlier identification. Upon applying the approach to the true J$1909$$-$$3744$ data from the NANOGrav $9$-year Dataset, we were able to identify several probable outliers that all belong to a single receiver--timing-backend system. Therefore, while this kind of outlier modeling can not necessarily reveal why contamination is occurring, it can diagnose its origins for further assessment. Finally, we deployed our approach on arguably the most prominent supermassive black-hole binary candidate found through periodicity searches in the Catalina Real-time Transient Survey: PG$1302$$-$$102$. We found that there can be significant variations in data quality across different surveys that are combined to make long baseline datasets, and appropriate care is needed to curate these for future binary analyses.

In summary, our outlier mitigation method aims to be versatile and easily implemented for different time-domain searches for supermassive black-hole binaries. Only small alterations in modeling are needed when adapting to each specific time-domain dataset, thereby providing a possible avenue to controlling any outlier contamination for multi-messenger time-domain studies. This practical method can assist with more robust statistical inference of model parameters, and give insight on the possible distribution and origin of outliers. The implementation as described here has already been incorporated into an outlier software package\footnote{\href{https://github.com/nanograv/enterprise_outliers}{https://github.com/nanograv/enterprise\_outliers}} that interfaces with the enterprise PTA analysis pipeline \citep{ellis_justin_a_2020_4059815}.

\section*{Acknowledgments}

We thank our colleagues in NANOGrav and the International Pulsar Timing Array for useful discussions about this work. We are particularly grateful for extended discussions with Justin~A.~Ellis and Michele Vallisneri, which improved the quality of this work. QW acknowledges the support of the Vanderbilt University Data Science Institute Summer Research Program (DSI-SRP). SRT acknowledges support from NSF AST-2007993, PHY-2020265, PHY-2146016, and a Vanderbilt University College of Arts \& Science Dean's Faculty Fellowship.  The Catalina Sky Survey is funded by NASA under Grant No. NNG05GF22G issued through the Science Mission Directorate Near-Earth Objects Observations Program. The CRTS survey is supported by the U.S. National Science Foundation under grants AST-0909182 and AST-1313422.

\section*{Data Availability}

The data underlying this article are available at \href{https://data.nanograv.org}{https://data.nanograv.org} for the NANOGrav $9$-year Dataset \citep{2015_NANOGrav9yr}, and \href{https://github.com/ZhuXJ1/SuperBayes}{https://github.com/ZhuXJ1/SuperBayes} for the CRTS, ASAS-SN, and LINEAR data combined in \citet{Zhu_2020}. Other data products are available upon request.



\bibliographystyle{mnras}
\bibliography{references} 

\begin{thebibliography}{}
\makeatletter
\relax
\def\mn@urlcharsother{\let\do\@makeother \do\$\do\&\do\#\do\^\do\_\do\%\do\~}
\def\mn@doi{\begingroup\mn@urlcharsother \@ifnextchar [ {\mn@doi@}
  {\mn@doi@[]}}
\def\mn@doi@[#1]#2{\def\@tempa{#1}\ifx\@tempa\@empty \href
  {http://dx.doi.org/#2} {doi:#2}\else \href {http://dx.doi.org/#2} {#1}\fi
  \endgroup}
\def\mn@eprint#1#2{\mn@eprint@#1:#2::\@nil}
\def\mn@eprint@arXiv#1{\href {http://arxiv.org/abs/#1} {{\tt arXiv:#1}}}
\def\mn@eprint@dblp#1{\href {http://dblp.uni-trier.de/rec/bibtex/#1.xml}
  {dblp:#1}}
\def\mn@eprint@#1:#2:#3:#4\@nil{\def\@tempa {#1}\def\@tempb {#2}\def\@tempc
  {#3}\ifx \@tempc \@empty \let \@tempc \@tempb \let \@tempb \@tempa \fi \ifx
  \@tempb \@empty \def\@tempb {arXiv}\fi \@ifundefined
  {mn@eprint@\@tempb}{\@tempb:\@tempc}{\expandafter \expandafter \csname
  mn@eprint@\@tempb\endcsname \expandafter{\@tempc}}}

\bibitem[\protect\citeauthoryear{Arzoumanian et~al.,}{Arzoumanian
  et~al.}{2015}]{2015_NANOGrav9yr}
Arzoumanian Z.,  et~al., 2015, \mn@doi [The Astrophysical Journal]
  {10.1088/0004-637x/813/1/65}, 813, 65

\bibitem[\protect\citeauthoryear{Arzoumanian et~al.,}{Arzoumanian
  et~al.}{2021}]{Arzoumanian_2021}
Arzoumanian Z.,  et~al., 2021, \mn@doi [The Astrophysical Journal]
  {10.3847/1538-4357/abfcd3}, 914, 121

\bibitem[\protect\citeauthoryear{{Begelman}, {Blandford}  \& {Rees}}{{Begelman}
  et~al.}{1980}]{1980Natur.287..307B}
{Begelman} M.~C.,  {Blandford} R.~D.,   {Rees} M.~J.,  1980, \mn@doi [\nat]
  {10.1038/287307a0}, \href
  {https://ui.adsabs.harvard.edu/abs/1980Natur.287..307B} {287, 307}

\bibitem[\protect\citeauthoryear{Betancourt}{Betancourt}{2018}]{betancourt2018conceptual}
Betancourt M.,  2018

\bibitem[\protect\citeauthoryear{{Burke-Spolaor} et~al.,}{{Burke-Spolaor}
  et~al.}{2019}]{2019A&ARv..27....5B}
{Burke-Spolaor} S.,  et~al., 2019, \mn@doi [\aapr] {10.1007/s00159-019-0115-7},
  \href {https://ui.adsabs.harvard.edu/abs/2019A&ARv..27....5B} {27, 5}

\bibitem[\protect\citeauthoryear{{Charisi}, {Bartos}, {Haiman}, {Price-Whelan}
  \& {M{\'a}rka}}{{Charisi} et~al.}{2015}]{2015MNRAS.454L..21C}
{Charisi} M.,  {Bartos} I.,  {Haiman} Z.,  {Price-Whelan} A.~M.,   {M{\'a}rka}
  S.,  2015, \mn@doi [\mnras] {10.1093/mnrasl/slv111}, \href
  {https://ui.adsabs.harvard.edu/abs/2015MNRAS.454L..21C} {454, L21}

\bibitem[\protect\citeauthoryear{{Charisi}, {Bartos}, {Haiman}, {Price-Whelan},
  {Graham}, {Bellm}, {Laher}  \& {M{\'a}rka}}{{Charisi}
  et~al.}{2016}]{2016Charisi}
{Charisi} M.,  {Bartos} I.,  {Haiman} Z.,  {Price-Whelan} A.~M.,  {Graham}
  M.~J.,  {Bellm} E.~C.,  {Laher} R.~R.,   {M{\'a}rka} S.,  2016, \mn@doi
  [\mnras] {10.1093/mnras/stw1838}, \href
  {https://ui.adsabs.harvard.edu/abs/2016MNRAS.463.2145C} {463, 2145}

\bibitem[\protect\citeauthoryear{{Charisi}, {Taylor}, {Runnoe}, {Bogdanovic}
  \& {Trump}}{{Charisi} et~al.}{2022}]{2022MNRAS.510.5929C}
{Charisi} M.,  {Taylor} S.~R.,  {Runnoe} J.,  {Bogdanovic} T.,   {Trump} J.~R.,
   2022, \mn@doi [\mnras] {10.1093/mnras/stab3713}, \href
  {https://ui.adsabs.harvard.edu/abs/2022MNRAS.510.5929C} {510, 5929}

\bibitem[\protect\citeauthoryear{Chen, Sesana  \& Conselice}{Chen
  et~al.}{2019}]{Chen_2019}
Chen S.,  Sesana A.,   Conselice C.~J.,  2019, \mn@doi [Monthly Notices of the
  Royal Astronomical Society] {10.1093/mnras/stz1722}, 488, 401–418

\bibitem[\protect\citeauthoryear{Chen et~al.,}{Chen et~al.}{2020}]{Chen_2020}
Chen Y.-C.,  et~al., 2020, \mn@doi [Monthly Notices of the Royal Astronomical
  Society] {10.1093/mnras/staa2957}

\bibitem[\protect\citeauthoryear{Chib \& Greenberg}{Chib \&
  Greenberg}{1995}]{ChibGreenberg}
Chib S.,  Greenberg E.,  1995, \mn@doi [The American Statistician]
  {10.1080/00031305.1995.10476177}, 49, 327

\bibitem[\protect\citeauthoryear{Colpi}{Colpi}{2014}]{Colpi_2014}
Colpi M.,  2014, \mn@doi [Space Science Reviews] {10.1007/s11214-014-0067-1},
  183, 189–221

\bibitem[\protect\citeauthoryear{Comerford, Pooley, Barrows, Greene, Zakamska,
  Madejski  \& Cooper}{Comerford et~al.}{2015}]{Comerford_2015}
Comerford J.~M.,  Pooley D.,  Barrows R.~S.,  Greene J.~E.,  Zakamska N.~L.,
  Madejski G.~M.,   Cooper M.~C.,  2015, \mn@doi [The Astrophysical Journal]
  {10.1088/0004-637x/806/2/219}, 806, 219

\bibitem[\protect\citeauthoryear{{D'Orazio}, {Haiman}  \&
  {Schiminovich}}{{D'Orazio} et~al.}{2015}]{2015Natur.525..351D}
{D'Orazio} D.~J.,  {Haiman} Z.,   {Schiminovich} D.,  2015, \mn@doi [\nat]
  {10.1038/nature15262}, \href
  {https://ui.adsabs.harvard.edu/abs/2015Natur.525..351D} {525, 351}

\bibitem[\protect\citeauthoryear{Desvignes et~al.,}{Desvignes
  et~al.}{2016}]{Desvignes_2016}
Desvignes G.,  et~al., 2016, \mn@doi [Monthly Notices of the Royal Astronomical
  Society] {10.1093/mnras/stw483}, 458, 3341–3380

\bibitem[\protect\citeauthoryear{{Drake} et~al.,}{{Drake}
  et~al.}{2009}]{2009ApJ...696..870D}
{Drake} A.~J.,  et~al., 2009, \mn@doi [\apj] {10.1088/0004-637X/696/1/870},
  \href {https://ui.adsabs.harvard.edu/abs/2009ApJ...696..870D} {696, 870}

\bibitem[\protect\citeauthoryear{Ellis, Vallisneri, Taylor  \& Baker}{Ellis
  et~al.}{2020}]{ellis_justin_a_2020_4059815}
Ellis J.~A.,  Vallisneri M.,  Taylor S.~R.,   Baker P.~T.,  2020, {ENTERPRISE:
  Enhanced Numerical Toolbox Enabling a Robust PulsaR Inference SuitE},
  \mn@doi{10.5281/zenodo.4059815}, \url
  {https://doi.org/10.5281/zenodo.4059815}

\bibitem[\protect\citeauthoryear{{Foster} \& {Backer}}{{Foster} \&
  {Backer}}{1990}]{1990ApJ...361..300F}
{Foster} R.~S.,  {Backer} D.~C.,  1990, \mn@doi [\apj] {10.1086/169195}, \href
  {https://ui.adsabs.harvard.edu/abs/1990ApJ...361..300F} {361, 300}

\bibitem[\protect\citeauthoryear{Gelfand}{Gelfand}{2000}]{Gelfand2000}
Gelfand A.~E.,  2000, Journal of the American Statistical Association, 95, 1300

\bibitem[\protect\citeauthoryear{Gelfand \& Smith}{Gelfand \&
  Smith}{1990}]{GelfandSmith1990}
Gelfand A.~E.,  Smith A. F.~M.,  1990, Journal of the American Statistical
  Association, 85, 398

\bibitem[\protect\citeauthoryear{Goicovic, Maureira-Fredes, Sesana,
  Amaro-Seoane  \& Cuadra}{Goicovic et~al.}{2018}]{Goicovic2018AccretionOC}
Goicovic F.,  Maureira-Fredes C.,  Sesana A.,  Amaro-Seoane P.,   Cuadra J.,
  2018, Monthly Notices of the Royal Astronomical Society, 479, 3438

\bibitem[\protect\citeauthoryear{Graham et~al.,}{Graham
  et~al.}{2015a}]{Graham_2015a}
Graham M.~J.,  et~al., 2015a, \mn@doi [Monthly Notices of the Royal
  Astronomical Society] {10.1093/mnras/stv1726}, 453, 1562–1576

\bibitem[\protect\citeauthoryear{Graham et~al.,}{Graham
  et~al.}{2015b}]{Graham_2015b}
Graham M.~J.,  et~al., 2015b, \mn@doi [Nature] {10.1038/nature14143}, 518,
  74–76

\bibitem[\protect\citeauthoryear{Haehnelt \& Kauffmann}{Haehnelt \&
  Kauffmann}{2002}]{HaehneltKauffmann2002}
Haehnelt M.~G.,  Kauffmann G.,  2002, \mn@doi [Monthly Notices of the Royal
  Astronomical Society] {10.1046/j.1365-8711.2002.06056.x}, 336, L61

\bibitem[\protect\citeauthoryear{Hogg, Bovy  \& Lang}{Hogg
  et~al.}{2010}]{hogg2010data}
Hogg D.~W.,  Bovy J.,   Lang D.,  2010, Data analysis recipes: Fitting a model
  to data (\mn@eprint {arXiv} {1008.4686})

\bibitem[\protect\citeauthoryear{Hopkins, Hernquist, Martini, Cox, Robertson,
  Di~Matteo  \& Springel}{Hopkins et~al.}{2005a}]{Hopkins_2005b}
Hopkins P.~F.,  Hernquist L.,  Martini P.,  Cox T.~J.,  Robertson B.,
  Di~Matteo T.,   Springel V.,  2005a, \mn@doi [The Astrophysical Journal]
  {10.1086/431146}, 625, L71–L74

\bibitem[\protect\citeauthoryear{Hopkins, Hernquist, Cox, Di~Matteo, Martini,
  Robertson  \& Springel}{Hopkins et~al.}{2005b}]{Hopkins_2005a}
Hopkins P.~F.,  Hernquist L.,  Cox T.~J.,  Di~Matteo T.,  Martini P.,
  Robertson B.,   Springel V.,  2005b, \mn@doi [The Astrophysical Journal]
  {10.1086/432438}, 630, 705–715

\bibitem[\protect\citeauthoryear{{Jayasinghe} et~al.,}{{Jayasinghe}
  et~al.}{2019}]{2019MNRAS.486.1907J}
{Jayasinghe} T.,  et~al., 2019, \mn@doi [\mnras] {10.1093/mnras/stz844}, \href
  {https://ui.adsabs.harvard.edu/abs/2019MNRAS.486.1907J} {486, 1907}

\bibitem[\protect\citeauthoryear{{Kelley}, {Blecha}, {Hernquist}, {Sesana}  \&
  {Taylor}}{{Kelley} et~al.}{2017}]{2017MNRAS.471.4508K}
{Kelley} L.~Z.,  {Blecha} L.,  {Hernquist} L.,  {Sesana} A.,   {Taylor} S.~R.,
  2017, \mn@doi [\mnras] {10.1093/mnras/stx1638}, \href
  {https://ui.adsabs.harvard.edu/abs/2017MNRAS.471.4508K} {471, 4508}

\bibitem[\protect\citeauthoryear{{Kelley}, {Blecha}, {Hernquist}, {Sesana}  \&
  {Taylor}}{{Kelley} et~al.}{2018}]{2018MNRAS.477..964K}
{Kelley} L.~Z.,  {Blecha} L.,  {Hernquist} L.,  {Sesana} A.,   {Taylor} S.~R.,
  2018, \mn@doi [\mnras] {10.1093/mnras/sty689}, \href
  {https://ui.adsabs.harvard.edu/abs/2018MNRAS.477..964K} {477, 964}

\bibitem[\protect\citeauthoryear{{Kelly}, {Bechtold}  \&
  {Siemiginowska}}{{Kelly} et~al.}{2009}]{2009ApJ...698..895K}
{Kelly} B.~C.,  {Bechtold} J.,   {Siemiginowska} A.,  2009, \mn@doi [\apj]
  {10.1088/0004-637X/698/1/895}, \href
  {https://ui.adsabs.harvard.edu/abs/2009ApJ...698..895K} {698, 895}

\bibitem[\protect\citeauthoryear{Kerr et~al.,}{Kerr et~al.}{2020}]{kerr_2020}
Kerr M.,  et~al., 2020, \mn@doi [Publications of the Astronomical Society of
  Australia] {10.1017/pasa.2020.11}, 37, e020

\bibitem[\protect\citeauthoryear{Khan, Fiacconi, Mayer, Berczik  \& Just}{Khan
  et~al.}{2016}]{Khan_2016}
Khan F.~M.,  Fiacconi D.,  Mayer L.,  Berczik P.,   Just A.,  2016, \mn@doi
  [The Astrophysical Journal] {10.3847/0004-637x/828/2/73}, 828, 73

\bibitem[\protect\citeauthoryear{{Kormendy} \& {Ho}}{{Kormendy} \&
  {Ho}}{2013}]{2013ARA&A..51..511K}
{Kormendy} J.,  {Ho} L.~C.,  2013, \mn@doi [\araa]
  {10.1146/annurev-astro-082708-101811}, \href
  {https://ui.adsabs.harvard.edu/abs/2013ARA&A..51..511K} {51, 511}

\bibitem[\protect\citeauthoryear{Liu et~al.,}{Liu et~al.}{2015}]{Liu_2015}
Liu T.,  et~al., 2015, \mn@doi [The Astrophysical Journal]
  {10.1088/2041-8205/803/2/l16}, 803, L16

\bibitem[\protect\citeauthoryear{Liu et~al.,}{Liu et~al.}{2016}]{Liu_2016}
Liu T.,  et~al., 2016, \mn@doi [The Astrophysical Journal]
  {10.3847/0004-637x/833/1/6}, 833, 6

\bibitem[\protect\citeauthoryear{{Liu}, {Gezari}  \& {Miller}}{{Liu}
  et~al.}{2018}]{2018ApJ...859L..12L}
{Liu} T.,  {Gezari} S.,   {Miller} M.~C.,  2018, \mn@doi [\apjl]
  {10.3847/2041-8213/aac2ed}, \href
  {https://ui.adsabs.harvard.edu/abs/2018ApJ...859L..12L} {859, L12}

\bibitem[\protect\citeauthoryear{Liu et~al.,}{Liu et~al.}{2019}]{Liu_2019}
Liu T.,  et~al., 2019, \mn@doi [The Astrophysical Journal]
  {10.3847/1538-4357/ab40cb}, 884, 36

\bibitem[\protect\citeauthoryear{{Lorimer} \& {Kramer}}{{Lorimer} \&
  {Kramer}}{2004}]{2004hpa..book.....L}
{Lorimer} D.~R.,  {Kramer} M.,  2004, {Handbook of Pulsar Astronomy}.
~ Vol. 4

\bibitem[\protect\citeauthoryear{{Milosavljevi{\'c}} \&
  {Merritt}}{{Milosavljevi{\'c}} \& {Merritt}}{2001}]{2001ApJ...563...34M}
{Milosavljevi{\'c}} M.,  {Merritt} D.,  2001, \mn@doi [\apj] {10.1086/323830},
  \href {https://ui.adsabs.harvard.edu/abs/2001ApJ...563...34M} {563, 34}

\bibitem[\protect\citeauthoryear{{Mingarelli} et~al.,}{{Mingarelli}
  et~al.}{2017}]{2017NatAs...1..886M}
{Mingarelli} C. M.~F.,  et~al., 2017, \mn@doi [Nature Astronomy]
  {10.1038/s41550-017-0299-6}, \href
  {https://ui.adsabs.harvard.edu/abs/2017NatAs...1..886M} {1, 886}

\bibitem[\protect\citeauthoryear{Mu{\~{n}}oz, Lai, Kratter  \&
  Miranda}{Mu{\~{n}}oz et~al.}{2020}]{Mu_oz_2020}
Mu{\~{n}}oz D.~J.,  Lai D.,  Kratter K.,   Miranda R.,  2020, \mn@doi [The
  Astrophysical Journal] {10.3847/1538-4357/ab5d33}, 889, 114

\bibitem[\protect\citeauthoryear{Neal}{Neal}{2011}]{Neal2011}
Neal R.~M.,  2011, Handbook of Markov Chain Monte Carlo.
Chapman and Hall/CRC, \mn@doi{10.1201/b10905}, \url
  {http://dx.doi.org/10.1201/b10905}

\bibitem[\protect\citeauthoryear{Perera et~al.,}{Perera
  et~al.}{2019}]{Perera_2019}
Perera B. B.~P.,  et~al., 2019, \mn@doi [Monthly Notices of the Royal
  Astronomical Society] {10.1093/mnras/stz2857}, 490, 4666–4687

\bibitem[\protect\citeauthoryear{{Pol} et~al.,}{{Pol}
  et~al.}{2021}]{2021ApJ...911L..34P}
{Pol} N.~S.,  et~al., 2021, \mn@doi [\apjl] {10.3847/2041-8213/abf2c9}, \href
  {https://ui.adsabs.harvard.edu/abs/2021ApJ...911L..34P} {911, L34}

\bibitem[\protect\citeauthoryear{{Ransom} et~al.,}{{Ransom}
  et~al.}{2019}]{2019BAAS...51g.195R}
{Ransom} S.,  et~al., 2019, in Bulletin of the American Astronomical Society.
  p.~195 (\mn@eprint {arXiv} {1908.05356})

\bibitem[\protect\citeauthoryear{Ritter \& Tanner}{Ritter \&
  Tanner}{1992}]{RitterTanner1992}
Ritter C.,  Tanner M.,  1992, \mn@doi [Journal of the American Statistical
  Association] {10.1080/01621459.1992.10475289}, 87, 861

\bibitem[\protect\citeauthoryear{{Rodriguez}, {Taylor}, {Zavala}, {Peck},
  {Pollack}  \& {Romani}}{{Rodriguez} et~al.}{2006}]{2006ApJ...646...49R}
{Rodriguez} C.,  {Taylor} G.~B.,  {Zavala} R.~T.,  {Peck} A.~B.,  {Pollack}
  L.~K.,   {Romani} R.~W.,  2006, \mn@doi [\apj] {10.1086/504825}, \href
  {https://ui.adsabs.harvard.edu/abs/2006ApJ...646...49R} {646, 49}

\bibitem[\protect\citeauthoryear{{Ryu}, {Perna}, {Haiman}, {Ostriker}  \&
  {Stone}}{{Ryu} et~al.}{2018}]{2018MNRAS.473.3410R}
{Ryu} T.,  {Perna} R.,  {Haiman} Z.,  {Ostriker} J.~P.,   {Stone} N.~C.,  2018,
  \mn@doi [\mnras] {10.1093/mnras/stx2524}, \href
  {https://ui.adsabs.harvard.edu/abs/2018MNRAS.473.3410R} {473, 3410}

\bibitem[\protect\citeauthoryear{{Sesana}, {Haiman}, {Kocsis}  \&
  {Kelley}}{{Sesana} et~al.}{2018}]{2018ApJ...856...42S}
{Sesana} A.,  {Haiman} Z.,  {Kocsis} B.,   {Kelley} L.~Z.,  2018, \mn@doi
  [\apj] {10.3847/1538-4357/aaad0f}, \href
  {https://ui.adsabs.harvard.edu/abs/2018ApJ...856...42S} {856, 42}

\bibitem[\protect\citeauthoryear{{Sesar}, {Stuart}, {Ivezi{\'c}}, {Morgan},
  {Becker}  \& {Wo{\'z}niak}}{{Sesar} et~al.}{2011}]{2011AJ....142..190S}
{Sesar} B.,  {Stuart} J.~S.,  {Ivezi{\'c}} {\v{Z}}.,  {Morgan} D.~P.,  {Becker}
  A.~C.,   {Wo{\'z}niak} P.,  2011, \mn@doi [\aj]
  {10.1088/0004-6256/142/6/190}, \href
  {https://ui.adsabs.harvard.edu/abs/2011AJ....142..190S} {142, 190}

\bibitem[\protect\citeauthoryear{Smith \& Gelfand}{Smith \&
  Gelfand}{1992}]{SmithGelfand1992_weighted}
Smith A. F.~M.,  Gelfand A.~E.,  1992, The American Statistician, 46, 84

\bibitem[\protect\citeauthoryear{Tak \& Morris}{Tak \& Morris}{2017}]{Tak_2017}
Tak H.,  Morris C.~N.,  2017, \mn@doi [Bayesian Analysis] {10.1214/16-ba1012},
  12

\bibitem[\protect\citeauthoryear{Tak, Ellis  \& Ghosh}{Tak
  et~al.}{2019}]{Ellis_2019}
Tak H.,  Ellis J.~A.,   Ghosh S.~K.,  2019, \mn@doi [Journal of Computational
  and Graphical Statistics] {10.1080/10618600.2018.1537925}, 28, 415

\bibitem[\protect\citeauthoryear{{Taylor}, {Vallisneri}, {Ellis}, {Mingarelli},
  {Lazio}  \& {van Haasteren}}{{Taylor} et~al.}{2016}]{2016ApJ...819L...6T}
{Taylor} S.~R.,  {Vallisneri} M.,  {Ellis} J.~A.,  {Mingarelli} C.~M.~F.,
  {Lazio} T.~J.~W.,   {van Haasteren} R.,  2016, \mn@doi [\apjl]
  {10.3847/2041-8205/819/1/L6}, \href
  {https://ui.adsabs.harvard.edu/abs/2016ApJ...819L...6T} {819, L6}

\bibitem[\protect\citeauthoryear{{Taylor}, {Simon}  \& {Sampson}}{{Taylor}
  et~al.}{2017}]{2017PhRvL.118r1102T}
{Taylor} S.~R.,  {Simon} J.,   {Sampson} L.,  2017, \mn@doi [\prl]
  {10.1103/PhysRevLett.118.181102}, \href
  {https://ui.adsabs.harvard.edu/abs/2017PhRvL.118r1102T} {118, 181102}

\bibitem[\protect\citeauthoryear{Vallisneri \& van Haasteren}{Vallisneri \& van
  Haasteren}{2017}]{Vallisneri_2017}
Vallisneri M.,  van Haasteren R.,  2017, \mn@doi [Monthly Notices of the Royal
  Astronomical Society] {10.1093/mnras/stx069}, p. stx069

\bibitem[\protect\citeauthoryear{{Vaughan}, {Uttley}, {Markowitz},
  {Huppenkothen}, {Middleton}, {Alston}, {Scargle}  \& {Farr}}{{Vaughan}
  et~al.}{2016}]{2016MNRAS.461.3145V}
{Vaughan} S.,  {Uttley} P.,  {Markowitz} A.~G.,  {Huppenkothen} D.,
  {Middleton} M.~J.,  {Alston} W.~N.,  {Scargle} J.~D.,   {Farr} W.~M.,  2016,
  \mn@doi [\mnras] {10.1093/mnras/stw1412}, \href
  {https://ui.adsabs.harvard.edu/abs/2016MNRAS.461.3145V} {461, 3145}

\bibitem[\protect\citeauthoryear{Verdinelli \& Wasserman}{Verdinelli \&
  Wasserman}{1991}]{Verdinelli1991BayesianAO}
Verdinelli I.,  Wasserman L.~A.,  1991, Statistics and Computing, 1, 105

\bibitem[\protect\citeauthoryear{{Witt}, {Charisi}, {Taylor}  \&
  {Burke-Spolaor}}{{Witt} et~al.}{2021}]{2021arXiv211007465W}
{Witt} C.~A.,  {Charisi} M.,  {Taylor} S.~R.,   {Burke-Spolaor} S.,  2021,
  arXiv e-prints, \href {https://ui.adsabs.harvard.edu/abs/2021arXiv211007465W}
  {p. arXiv:2110.07465}

\bibitem[\protect\citeauthoryear{{Xin}, {Charisi}, {Haiman}, {Schiminovich},
  {Graham}, {Stern}  \& {D'Orazio}}{{Xin} et~al.}{2020}]{2020MNRAS.496.1683X}
{Xin} C.,  {Charisi} M.,  {Haiman} Z.,  {Schiminovich} D.,  {Graham} M.~J.,
  {Stern} D.,   {D'Orazio} D.~J.,  2020, \mn@doi [\mnras]
  {10.1093/mnras/staa1643}, \href
  {https://ui.adsabs.harvard.edu/abs/2020MNRAS.496.1683X} {496, 1683}

\bibitem[\protect\citeauthoryear{Yu}{Yu}{2002}]{yu2002}
Yu Q.,  2002, \mn@doi [Monthly Notices of the Royal Astronomical Society]
  {10.1046/j.1365-8711.2002.05242.x}, 331, 935

\bibitem[\protect\citeauthoryear{Zhu \& Thrane}{Zhu \& Thrane}{2020}]{Zhu_2020}
Zhu X.-J.,  Thrane E.,  2020, \mn@doi [The Astrophysical Journal]
  {10.3847/1538-4357/abac5a}, 900, 117

\makeatother
\end{thebibliography}





\bsp	
\label{lastpage}
\end{document}